%% file: paper.tex
\RequirePackage{silence}
\documentclass[letterpaper,twocolumn,10pt]{article}
\usepackage{usenix}
\usepackage[available,functional,reproduced]{usenixbadges}
\usepackage[utf8]{inputenc}
\microtypecontext{spacing=nonfrench}

\usepackage{tikz}
\usepackage{amsmath}
\usepackage{amssymb}
\usepackage{pifont}
\usepackage{xspace}
\usepackage{booktabs}
\usepackage{subcaption}
\usepackage{bytefield}
\usepackage{rotating}
\usepackage[noabbrev]{cleveref}
\usepackage{tablefootnote}
\usepackage{soul}
\usepackage{csquotes}
\usepackage{wasysym}
\usepackage{multirow}
\usepackage{paralist}
\usepackage[inline]{enumitem}

\usepackage{threeparttable}
\usepackage{tabularx}
\usepackage{listings}

\let\oldparagraph=\paragraph
\renewcommand\paragraph[1]{\oldparagraph{#1.}}

\newcommand{\artifactsUrl}{
\url{https://github.com/RUB-NDS/Terrapin-Artifacts}
}

\newcommand{\mitm}{MitM\xspace}
\newcommand{\sshmsg}[1]{\textsc{#1}\xspace}
\newcommand{\msgkexinit}{\sshmsg{KexInit}}
\newcommand{\msgnewkeys}{\sshmsg{NewKeys}}
\newcommand{\msgkexdhinit}{\sshmsg{KexDhInit}}
\newcommand{\msgkexdhreply}{\sshmsg{KexDhReply}}
\newcommand{\msgservicerequest}{\sshmsg{ServiceRequest}}
\newcommand{\msgserviceaccept}{\sshmsg{ServiceAccept}}
\newcommand{\msgignore}{\sshmsg{Ignore}}
\newcommand{\msgunknown}{\sshmsg{Unknown}}
\newcommand{\msgunimplemented}{\sshmsg{Unimplemented}}
\newcommand{\msgextinfo}{\sshmsg{ExtInfo}}
\newcommand{\msguserauthrequest}{\sshmsg{UserAuthRequest}}
\newcommand{\msgping}{\sshmsg{Ping}}
\newcommand{\msgpong}{\sshmsg{Pong}}
\newcommand{\msgdisconnect}{\sshmsg{Disconnect}}

\newcommand{\ivkdf}{\ensuremath{\mathrm{IV}_\mathrm{KDF}}\xspace}

\newcommand{\scanSupported}{71.6}
\newcommand{\scanPreferred}{63.2}

\newcommand{\seqno}[1]{\ensuremath{\mathsf{#1}}\xspace}
\newcommand{\csnd}{\seqno{C.Snd}}
\newcommand{\crcv}{\seqno{C.Rcv}}
\newcommand{\ssnd}{\seqno{S.Snd}}
\newcommand{\srcv}{\seqno{S.Rcv}}
\newcommand{\xsnd}{\seqno{Snd}}
\newcommand{\xrcv}{\seqno{Rcv}}
\newcommand{\atkrcvinc}{RcvIncrease\xspace}
\newcommand{\atkrcvdec}{RcvDecrease\xspace}
\newcommand{\atksndinc}{SndIncrease\xspace}
\newcommand{\atksnddec}{SndDecrease\xspace}

\date{}

\title{\Large \bf Terrapin Attack: Breaking SSH Channel Integrity \\By Sequence Number Manipulation}

\author{
{\rm Fabian Bäumer}\\
Ruhr University Bochum
\and
{\rm Marcus Brinkmann}\\
Ruhr University Bochum
\and
{\rm Jörg Schwenk}\\
Ruhr University Bochum
}

\begin{document}

\maketitle

\begin{abstract}
\input{sections/000-abstract}
\end{abstract}

\section{Introduction}
\input{sections/010-introduction}

\section{Related Work}
\input{sections/020-related-work}

\section{Background}
\input{sections/030-background}

\section{Breaking SSH Channel Integrity}
\input{sections/050-attacks}

\section{Breaking SSH Extension Negotiation}
\input{sections/055-rw-attacks}

\section{Message Injection Attacks on AsyncSSH}
\input{sections/057-asyncssh}

\section{SSH Deployment Statistics}
\input{sections/065-statistics}

\section{Suggested Countermeasures}
\input{sections/070-countermeasures}

\section{Future Work}
\input{sections/075-future}

\section{Conclusion}
\input{sections/080-conclusion}

\paragraph{Acknowledgements}
\input{sections/090-ack}

\bibliographystyle{plain}
\bibliography{paper,abbrev3,crypto,rfc}


\newpage
\renewcommand\paragraph[1]{\oldparagraph{#1}}
\input{ae/appendix.tex}

\end{document}

%% file: sections/000-abstract.tex

The SSH protocol provides secure access to network services, particularly
remote terminal login and file transfer within organizational networks and to
over 15~million servers on the open internet. SSH uses an authenticated key
exchange to establish a \emph{secure channel} between a client and a server,
which protects the confidentiality and integrity of messages sent in either
direction. The secure channel prevents message manipulation, replay, insertion,
deletion, and reordering. At the network level, SSH uses the \emph{Binary
Packet Protocol} over TCP.

In this paper, we show that as new encryption algorithms and mitigations were
added to SSH, the SSH Binary Packet Protocol is no longer a secure channel: SSH
channel integrity (INT-PST, aINT-PTXT, and INT-sfCTF) is broken for three
widely used encryption modes. This allows \emph{prefix truncation attacks}
where encrypted packets at the beginning of the SSH channel can be deleted
without the client or server noticing it. We demonstrate several real-world
applications of this attack. We show that we can fully break SSH extension
negotiation (RFC~8308), such that an attacker can downgrade the public key
algorithms for user authentication or turn off a new countermeasure against
keystroke timing attacks introduced in OpenSSH~9.5. Further, we identify an
implementation flaw in AsyncSSH that, together with prefix truncation, allows
an attacker to redirect the victim's login into a shell controlled by the
attacker.

We also performed an internet-wide scan for affected encryption modes and
support for extension negotiation. We find that \scanSupported\% of SSH servers
support a vulnerable encryption mode, while \scanPreferred\% even list it as
their preferred choice.

We identify two root causes that enable these attacks: First, the SSH handshake
supports optional messages that are not authenticated. Second, SSH does not
reset message sequence numbers when activating encryption keys. Based on this
analysis, we propose effective and backward-compatible changes to SSH that
mitigate our attacks.

%% file: sections/010-introduction.tex

\paragraph{Secure Shell (SSH)}
While TLS is commonly used to secure user-facing protocols such as web, email,
or FTP, SSH is used by administrators to deploy and maintain these servers,
often with high privilege (root) access and a large attack surface for lateral
movement within an organization's infrastructure. SSH was developed by Tatu
Ylonen in 1995 as a secure alternative to telnet and rlogin/rcp and has since
become a critical component of internet security.

In 1996, SSHv2 was developed to fix severe vulnerabilities in the original
version. In February~1997, the IETF formed the SECSH working group to
standardize SSHv2. After a decade, it published five core RFCs~\cite{RFC4250,
RFC4251,RFC4252,RFC4253,RFC4254}.  SSHv2 provides cryptographic agility and
protocol agility without breaking backward compatibility. Since its original
release, dozens of standardized and informal updates to the protocol have been
published. Because of this, SSHv2 remains relevant after 25~years without major
redesign, but it has also become difficult to analyze. There is a significant
risk that these extensions of SSH interact to undermine its security goals.

\paragraph{SSH Connections}
An SSH connection between a client and a server begins with the \emph{Transport
Layer Protocol}~\cite{RFC4253}, which defines the handshake messages for key
exchange and server authentication and how messages are exchanged over TCP
using the \emph{Binary Packet Protocol} (BPP). After the handshake, SSH
provides a \emph{secure channel}\footnote{We note that by \emph{channel}, we
refer to the integrity-protected, encrypted byte stream at the transport level,
not the SSH data channels from the connection protocol. Multiple SSH data
channels can be multiplexed over the same secure transport channel.} for
application data. At the application level, the client chooses a sequence of
\emph{services} to run. In practice, the client will run precisely two
services: the \emph{Authentication Protocol}~\cite{RFC4252} for user
authentication with a password or public key, followed by the \emph{Connection
Protocol}~\cite{RFC4254} for the bulk of SSH's features like terminal sessions,
port forwarding, and file transfer.

\subsection{SSH Channel Security}

\begin{figure}[t]
    \centering
    \frame{\includegraphics[width=\columnwidth]{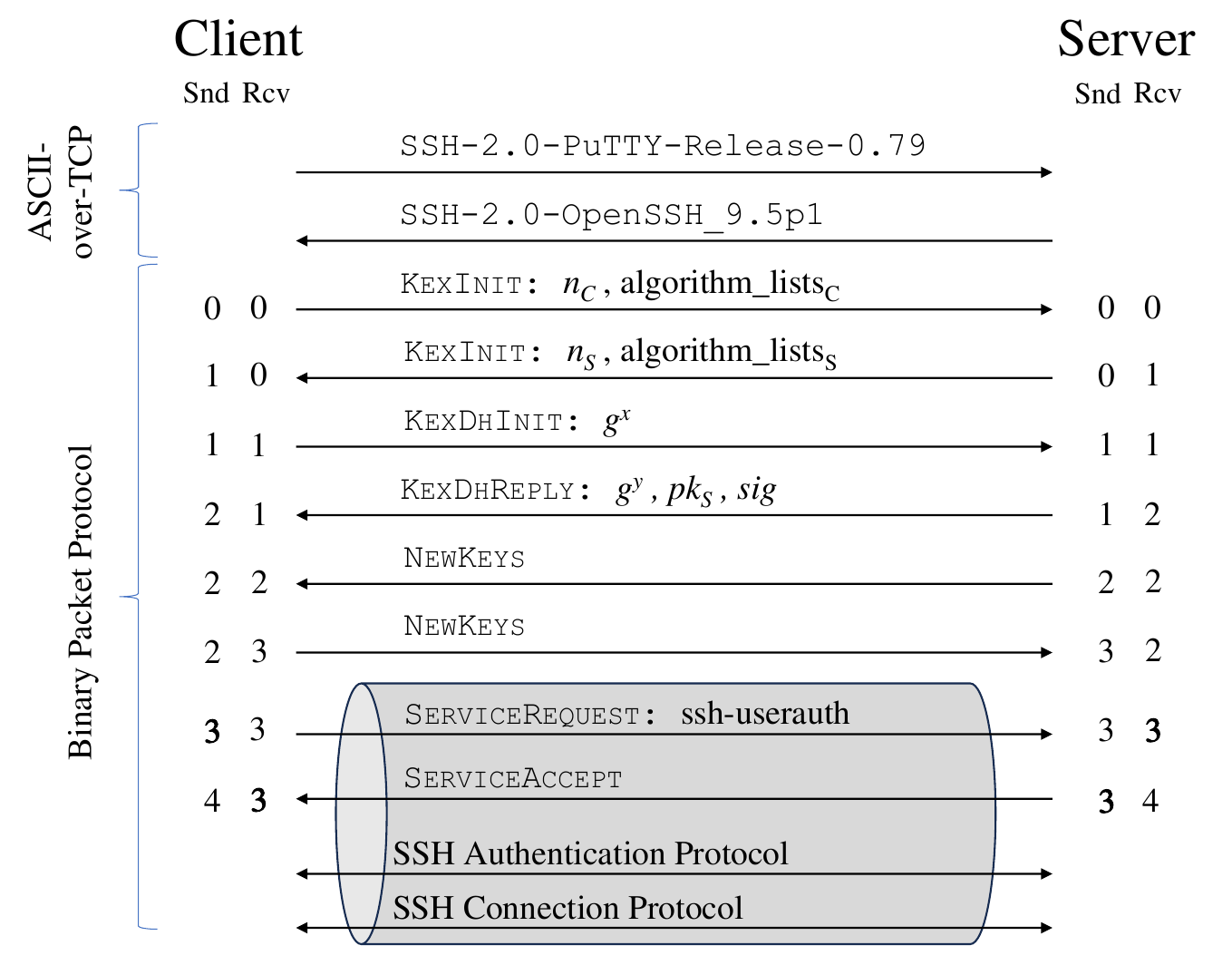}}
\caption{Typical SSH handshake using a finite-field Diffie-Hellman key exchange. Included sequence numbers are implicit and maintained by the BPP. \xsnd denotes the counter for sent packets and \xrcv for received packets. Sequence numbers verified using authenticated encryption are in \textbf{bold}.}
    \label{fig:ssh-hs}
\end{figure}

In this work, we focus on the integrity of the SSH handshake and the resulting
secure channel, as shown in~\autoref{fig:ssh-hs}. After an initial exchange of
version information directly over TCP, the BPP exchanges \emph{packets}, each
containing precisely one \emph{message}. Initially, the BPP is used without
encryption or authentication for the duration of the handshake until the
\msgnewkeys message. Afterward, the encryption and authentication keys are used
to form a secure channel, intending to protect the confidentiality and
integrity of the \emph{ordered stream of all following messages}. Note that
technically, the secure channel consists of two separate cipher streams, one
for each direction, and that the order of message arrival is only guaranteed
for each direction separately.

\paragraph{Message Authentication Codes}
As SSH is an interactive protocol, the integrity of each packet must be
verified when it is received so that it can be promptly processed. For this,
the BPP appends a Message Authentication Code (MAC) to each packet. A cipher
mode and a MAC form an authenticated encryption scheme~\cite{AC:BelNam00}. SSH
historically uses \emph{Encrypt-and-MAC} (EaM), where the MAC is computed over
the plaintext, but this is vulnerable to oracle attacks~\cite{SP:AlbPatWat09}.
Later, \emph{Encrypt-then-MAC} (EtM) was added, where the MAC is computed over
the ciphertext instead. SSH has recently adopted the AEAD ciphers AES-GCM and
ChaCha20-Poly1305, where ciphertext integrity is built into the encryption
scheme~\cite{CCS:Rogaway02}.

\paragraph{A Trivial Example: Suffix Truncation Attacks}
Note that a per-packet MAC cannot fully protect the channel's integrity, as
packets are verified and decrypted before the end of communication has been
seen. This allows for a trivial \emph{suffix truncation attack}, where the
attacker interrupts the message flow at some point during the communication.
This is an inherent limitation of interactive protocols and an accepted
trade-off in the design of SSH, but also, e.g., the TLS Record Layer. Although
this attack cannot be prevented, it can be detected by requiring \enquote
{end-of-communication} messages as the last messages in both directions. TLS
defines a \enquote{close\_notify} alert for this purpose~\cite{RFC8446}.
Although SSH also defines a \msgdisconnect message to indicate the end of the
secure channel, this message is optional, unidirectional, and not described as
security-critical in the standard.

\paragraph{Implicit Sequence Numbers}
If the MAC was only computed over the payload of each packet, an attacker could
still delete, replay, or reorder packets. Therefore, a \emph{sequence number}
is included in the MAC computation, corresponding to the position of the
message in the stream. Each peer maintains two counters (starting at~0), one
for each direction. The \xsnd counter is incremented after a packet has been
sent, and the \xrcv counter is incremented after a received packet has been
processed. Once the secure channel has been established, the current value of
\xsnd is used to compute the MAC of an outgoing packet, and the current value
of \xrcv is used to verify the MAC of an incoming packet. If packets in the
secure channel are deleted, replayed, or reordered, the sequence numbers get
out of sync, and MAC verification will fail.

Because TCP is a reliable transport, accidental reordering of SSH packets
cannot occur on the network. Thus, SSH (like other TCP-based protocols) uses
\emph{implicit sequence numbers} that are not transmitted as part of the packet.

\paragraph{Security Guarantees of Secure Channels}
For TLS, the security guarantees of the Record Layer were formalized as
stateful length-hiding encryption \cite{AC:PatRisShr11}, with the state mainly
consisting of the implicit sequence number. The security of the BPP and
implicit sequence numbers was analyzed by Bellare et al. in
\cite{CCS:BelKohNam02} and later refined and extended by Paterson and Watson
\cite{EC:PatWat10} and Albrecht et al.~\cite{CCS:ADHP16}. These works define,
in slightly idealized scenarios, the following informal security goal for a
secure channel:
\begin{quote}
    When a secure channel between A and B is used, the data (or message) stream
    received by B should be identical to the one sent by A and vice versa
    (INT-PST, aINT-PTXT in~\cite{C:FGMP15}).
\end{quote}
Within their idealizations, all three works confirm that the BPP is indeed a
secure channel. The difference between the models is that Paterson and
Watson~\cite{EC:PatWat10} also included the encrypted length field of the
Encrypt-and-MAC modes, while Albrecht et al.~\cite{CCS:ADHP16} considered the
more recent cipher modes ChaCha20-Poly1305, AES-GCM, and generic
Encrypt-then-MAC. Our attacks show that the models underlying the proofs
in~\cite{CCS:ADHP16} are only partially accurate. We will explain the
discrepancies between the proofs and our findings
in~\autoref{sec:whyproofswrong}.

\subsection{Overview of Our Attacks on SSH}
\begin{figure}[t]
    \centering
    \frame{\includegraphics[width=\columnwidth]{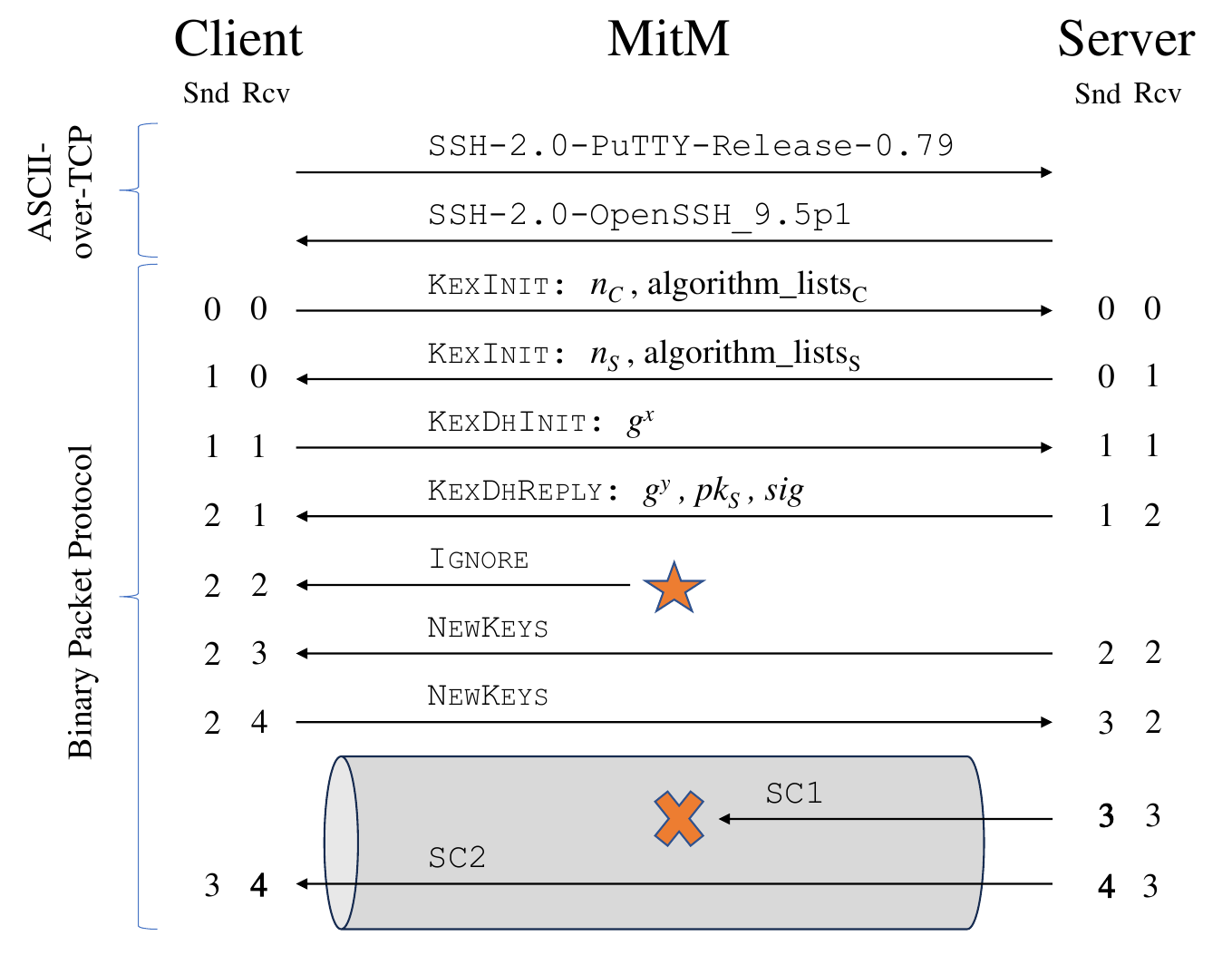}}
    \caption{A novel prefix truncation attack on the BPP. The server sends
        \textsf{SC1} and \textsf{SC2}, but the client only receives \textsf
        {SC2}.}
    \label{fig:ssh-attack}
\end{figure}

In this paper, we show that SSH fails to protect the integrity of the encrypted
message stream against meddler-in-the-middle (\mitm) attacks. More precisely,
we present novel \emph{prefix truncation attacks} against SSH:
\begin{quote}
We show that the SSH Binary Packet Protocol is \emph{not} a secure channel
because a \mitm attacker can delete a chosen number of integrity-protected
packets from the beginning of the channel in either or both directions without
being detected (\autoref{fig:ssh-attack}).
\end{quote}

\paragraph{Attacker Model}
We consider a \mitm attacker who can observe, change, delete, or insert bytes
at the TCP layer. We do \emph{not} assume that the attacker can break the
confidentiality of the session keys, i.e., the attacker has no information
about the derived encryption keys, MAC keys, or IV. However, we do assume that
the attacker can determine the length of the messages to be deleted even if the
length field is encrypted. We discuss the practicality of this in
\autoref{sec:ssh-attack:par:bytelength}. The rogue session attack presented in
\autoref{sec:asyncrogueshell} further assumes the attacker has an account on
the same host as the victim.

As for the connection, we assume that the server is correctly authenticated
(i.e., the client recognizes the server's host key) and that a vulnerable
encryption mode has been negotiated. See \autoref{tab:encryptionmodes} for a
list of vulnerable encryption modes.

\paragraph{Prefix Truncation Attacks}
While our attacks on SSH are novel, the idea of prefix truncation attacks
against network protocols by sequence number manipulation is not. To the best
of our knowledge, the first and only description of such an attack is by
Fournet (on behalf of miTLS) in an email to the TLS working group in 2015,
targeting a draft version of TLS~1.3~\cite{fournetmail}. Fournet's attack
increases sequence numbers in TLS by message fragmentation rather than message
injection and remains theoretical, as \emph{``prefix truncations will probably
cause the handshake to fail.''} Subsequently, the draft was modified, and no
prefix truncation attacks against the final version of TLS~1.3 are known. In
contrast, we present the first real-world, practical prefix truncation attack
against a mature, widely used protocol.

\paragraph{Root Cause Analysis}
Our results depend on two technical observations about how SSH protects the
integrity of the handshake and channel:
\begin{enumerate}
    \item \emph{SSH does not protect the full handshake transcript.} Although
    server authentication uses a signature to verify the integrity of the
    handshake, the signature is formed over a fixed list of handshake messages
    rather than the complete transcript. This gap in authentication allows an
    attacker to insert messages into the handshake and thereby manipulate
    sequence numbers.

    \item \emph{SSH does not reset sequence numbers at the beginning of the
    secure channel.} Instead, SSH increases sequence numbers monotonically,
    independent of the encryption state. Any manipulation of sequence numbers
    before the secure channel carries over into the channel.
\end{enumerate}
Based on these two key observations, we present a series of novel attacks on SSH that increase in complexity and impact.

\paragraph{Sequence Number Manipulation}
We show that an attacker can \emph{increase the receive counters} of the server
and the client by inserting messages into the handshake. Although not required
for any of our attacks, we also show that, for some implementations, an
attacker can \emph{fully control the receive and send counters}, setting them
to arbitrary values (\autoref{sec:advancedsqnnomanipulation}).

\paragraph{A Prefix Truncation Attack on the BPP}
An attacker can use sequence number manipulation to \emph{delete a chosen
number of packets at the beginning of the secure channel}. Neither the client
nor the server detects this prefix truncation, consequently breaking the
channel integrity of SSH (\autoref{sec:ssh-attack}).

\paragraph{Extension Negotiation Downgrade Attack}
As a practical example, we show an attack that uses prefix truncation to
\emph{break extension negotiation} \cite{RFC8308}, thereby downgrading the
security of the connection. The attacked client might mistakenly believe that
the server does not support recent signature algorithms for user authentication
or does not implement certain countermeasures to attacks
(\autoref{sec:extdowngradeattack}). Specifically, the attacker can turn off
protection against keystroke timing attacks in the recently released
OpenSSH~9.5.

\paragraph{Rogue Extension Attack and Rogue Session Attack}
As another example, we show two attacks on the AsyncSSH client and server. In
the first attack, \emph{the victim's extension info message is replaced} with
one chosen by the attacker (\autoref{sec:asyncrogueext}). For the second
attack, the attacker must have a user account on the same server as the victim.
The attacker injects a malicious user authentication message so that \emph{the
victim logs into a shell controlled by the attacker} rather than the victim's
shell, thereby giving the attacker complete control over the victim's terminal
input (\autoref{sec:asyncrogueshell}). These attacks combine prefix truncation
with implementation flaws in the AsyncSSH library.

\paragraph{Limitations}
Our attacks critically depend on the SSH encryption mode negotiated between the
client and the server. The attack works best with the AEAD cipher
ChaCha20-Poly1305 (added in 2013). The attack also works with any EtM mode
(added in 2012), although the success probability depends on the cipher mode
negotiated. CBC-EtM can be exploited with a significant probability, while the
exploitability of CTR-EtM is low. On the other hand, CBC-EaM, CTR-EaM, and GCM
modes are not affected. See~\autoref{sec:vulnciphers} for a complete analysis.

In an internet-wide scan, we show that despite these limitations,
\scanSupported\% of all SSH servers on the internet support an affected
encryption mode, and \scanPreferred\% even list it as their preferred choice
(\autoref{sec:internetscan}).

\subsection{Our Contributions}
We contribute the following novel results:

\begin{itemize}
    \item An analysis of the integrity of SSH channels, where we identify two
    previously unknown flaws in the SSH specification, namely gaps in the
    handshake authentication and the use of sequence numbers across key
    activation.

    \item A novel prefix truncation attack on SSH channel integrity, where we
    show that an attacker can manipulate the sequence numbers and delete
    several messages from the beginning of the secure channel.

    \item A first security analysis of SSH extension negotiation, including a
    novel downgrade attack that disables extension negotiation completely.
    Thus, support for some public key signature algorithms or, with
    OpenSSH~9.5, protection against keystroke timing attacks can be disabled.

    \item As a practical demonstration, two novel attacks on AsyncSSH. First, a
    rogue extension attack, where the attacker can insert a chosen extension
    negotiation message. Second, a rogue session attack that allows the
    attacker to log the victim into an attacker-controlled shell. Both escalate
    implementation flaws in AsyncSSH using the prefix truncation attack.

    \item An internet-wide scan with up-to-date information on the distribution
    of SSH encryption modes and extensions.
\end{itemize}

\paragraph{Artifacts}
Proof-of-concept implementations for our attacks and the aggregated results of
our internet-wide scan are available under the Apache-2.0 open-source license.
See:

\artifactsUrl

\paragraph{Ethics Consideration and Responsible Disclosure}
\label{sec:ethicalconsiderations}
We disclosed our findings to 33 vendors of SSH implementations, including
OpenSSH and AsyncSSH, in October and November 2023, followed by a public
disclosure on December 18th, 2023. As of February 2024, 28 vendors have
published patches implementing a backward-compatible countermeasure proposed by
OpenSSH. The general protocol flaw has been assigned CVE-2023-48795
(CVSSv3 5.9), while the implementation flaws in AsyncSSH were assigned
CVE-2023-46445 (Rogue Extension Negotiation; CVSSv3 5.9) and CVE-2023-46446
(Rogue Session Attack; CVSSv3 6.8). To estimate the adoption rate of the
countermeasure, we scanned the IPv4 address space on January 5th, 2024,
indicating that more than 3.4M servers were patched.

We provide an opt-out option and an email address for inquiries about our
internet-wide scans. Additionally, we employ a block list to exclude networks
that opted out of previous scans. Scan results are solely published in
aggregated form, without any information that could identify individual servers
or networks.

%% file: sections/020-related-work.tex

\paragraph{Secure Channels}
In 2001, Canetti and Krawczyk \cite{EC:CanKra01} established the first model
for secure channels, which only requires protection against adversarial
insertion of messages. Paterson et al. \cite{AC:PatRisShr11} defined stateful
length-hiding authenticated encryption (sLHAE) to model the TLS record layer as
a secure channel. This definition was used in \cite{C:JKSS12,C:KraPatWee13} to
define authenticated and confidential channel establishment (ACCE) to analyze
the TLS handshake and record layer as a whole.
Bellare et al.~\cite{CCS:BelKohNam02} used stateful authenticated decryption to
define a security notion for SSH that is directed against replay and
out-of-order delivery attacks (INT-sfCTXT). Paterson and
Watson~\cite{EC:PatWat10} later refined this work to cover buffered decryption
(INT-BSF-CTXT). Albrecht et al.~\cite{CCS:ADHP16} further refined and extended
this definition to cover ciphertext fragmentation attacks more generally
(INT-sfCTF). Generalizing the work on TLS and SSH,
Fishlin et al.~\cite{C:FGMP15} defined, among other notions,
plaintext integrity for generic data and (atomic) message streams (INT-PST,
aINT-PTXT).

Our attacks show that SSH BPP, when instantiated with ChaCha20-Poly1305,
CBC-EtM, or CTR-EtM, does not provide integrity of plaintext or ciphertext
(message) streams (INT-PST, aINT-PST, INT-sfCTF) as defined
in~\cite{CCS:ADHP16,C:FGMP15}.

\paragraph{Truncation Attacks}
Suffix truncation attacks against web services using TLS have been demonstrated
by Smyth and Pironti in~\cite{smyth2013}. A prefix truncation attack against a
draft version of TLS~1.3 was described by Fournet (on behalf of miTLS) in an
email to the TLS working group in 2015~\cite{fournetmail}. Fournet's attack
increases TLS sequence numbers by message fragmentation rather than injection
to avoid breaking handshake authentication. The attack remained theoretical as
\emph{``prefix truncations will probably cause the handshake to fail.''} As a
countermeasure, the draft was changed back to reset sequence numbers to~0 when
activating keys.

\paragraph{Attacks on SSH}
The most severe attack on SSH was presented by Albrecht, Paterson, and
Watson~\cite{SP:AlbPatWat09} in 2009. It exploited the encrypted length field,
using the length of the ciphertext accepted by the server from the network as a
decryption oracle for parts of a ciphertext block. In \cite{EC:PatWat10}, this
peculiarity of the BPP was formalized, and in \cite{CCS:ADHP16} a variant of
this attack was presented. Other attacks on SSH include a timing attack on SSH
keystrokes by Song, Wagner, and Tian~\cite{USENIX:SonWagTia01}, a theoretical
attack on SSH CBC cipher modes by Wei Dai~\cite{weidai2002}, and a SHA-1 chosen
prefix collision attack on the handshake transcript by Bhargavan and
Leurent~\cite{NDSS:BhaLeu16}. The weakness of some SSH host keys presented by
Heninger et al.~\cite{USENIX:HDWH12} was caused by a lack of entropy and faulty
implementations and is not an inherent weakness of the protocol.

\paragraph{Formal Proofs for SSH}
\label{sec:whyproofswrong}
The SSH handshake was analyzed by Williams~\cite{IMA:Williams11} and
Bergsma et al.~\cite{CCS:BDKSS14}. Bellare et al.~\cite{CCS:BelKohNam02}
presented a generic security model for SSH BPP, and Paterson and
Watson~\cite{EC:PatWat10} a specific, more detailed one for CTR-EaM.
Albrecht et al.~\cite{CCS:ADHP16} included security statements for
ChaCha20-Poly1305, generic Encrypt-then-MAC, and AES-CTR in SSH, claiming the
indistinguishability and integrity of the ciphertext. Careful analysis of their
proofs reveals an essential assumption about SSH sequence numbers that does not
hold. In particular, they assume that the sequence counters in the stateful
encryption scheme are initialized to~0 on both sides, which is false for the
cipher modes affected by our attack. This assumption is not apparent from the
paper, which omits the pseudocode for the encryption schemes, but Hansen gives
the missing parts in~\cite{Hansen2020} (Alg.~\verb|ssh-ChaCha20-Poly1305-Gen|
in Fig.~6.5 and Alg.~\verb|ssh-fgEtM-Gen| in Fig.~6.6 there). We note that this
assumption is also present in~\cite{CCS:BelKohNam02} (Fig.~4 there).

Cadé and Blanchet~\cite{cryptoverifssh} used the formal verification tool
\emph{CryptoVerif}~\cite{C:BlaPoi06} to prove the security of SSH server
authentication and the secrecy of the session key in the computational model.
The secrecy of messages in the channel cannot be shown due to the attack
in~\cite{SP:AlbPatWat09}. However, they do mention that due to a limitation in
the design of CryptoVerif, it cannot keep mutable internal states such as
sequence numbers or counters. In their model, the sequence numbers are passed
explicitly as arguments and are, therefore, under the attacker's control. The
authors do not raise the issue of channel integrity. Other computer-aided
proofs of server authentication and secrecy of the session key in the symbolic
or computational model can be found in~\cite{lafourcade2016,cheval2022}, which
also do not consider the integrity of the secure channel. For an overview of
the field of computer-aided cryptography, see~\cite{SP:BBBBCLP21}.

%% file: sections/030-background.tex

\label{sec:background}

\paragraph{SSH Handshake (\autoref{fig:ssh-hs})}
To initiate an SSH connection, both peers exchange a version banner. The Binary
Packet Protocol (see below) is used from the third message on but without
encryption and authentication. In the \msgkexinit messages, nonces and ordered
lists of algorithms are exchanged: One list for key exchange, one for server
signatures, and two (one per direction) each for encryption, MAC, and
compression. For each list, the negotiated algorithm is the first algorithm in
the client's list, which is also offered by the server.

In the \msgkexdhinit and \msgkexdhreply messages, a finite-field Diffie-Hellman
key exchange is performed. SSH also supports elliptic curves (ECDH) and hybrid
schemes with post-quantum cryptography (PQC) as alternatives. The server
authenticates itself with a digital signature as part of the handshake. The
signature is computed over the contents of the previously exchanged messages in
a specified order.

\paragraph{The Exchange Hash: A Partial Handshake Transcript}
\label{par:exchangehash}
In contrast to TLS, SSH uses only a selection from the handshake transcript for
authentication. The hash value computed from this selection is called
\emph{exchange hash} $H$, defined as
\[ H = \text{HASH}(V_C~||~V_S~||~I_C~||~I_S~||~K_S~||~X~||~K), \] where HASH is
the hash function of the negotiated key exchange, $V_C$ and $V_S$ are the
version banners of the client and server, $I_C$ and $I_S$ are the \msgkexinit
messages, $K_S$ is the server's public host key, and $K$ is the shared secret
derived from the key exchange. The value of $X$ depends on the key exchange and
contains a composition of negotiated parameters (if any) and the ephemeral
public keys of the key exchange~\cite[Sec. 8]{RFC4253}. Each field includes a
length field defined by the encoding.

Although the exchange hash contains everything that may influence the
negotiation of algorithms or computation of the shared secret, it excludes
seemingly `unimportant' messages or message parts, such as \msgignore messages
and unrecognized messages. This authentication gap allows a \mitm attacker to
inject messages into the handshake.

\paragraph{Sequence Numbers}
Each sequence number is stored as a 4-byte unsigned integer initialized to zero
upon connection. After a binary packet has been sent or received, the
corresponding sequence number \xsnd or \xrcv is incremented by one. Sequence
numbers are never reset for a connection but roll over to~0 after $2^{32}-1$.
As sequence numbers are responsible for protecting against replay attacks,
rekeying must occur at least once every $2^{32}$
packets~\cite[Sec. 6.1]{RFC4344}.

We illustrate the use of sequence numbers in~\autoref{fig:ssh-hs}: After the
banner exchange, the counters \xsnd and \xrcv are initialized with $(0,0)$ on
both sides. During algorithm negotiation and key exchange, sequence numbers are
increased but not used in any MAC computation or verification. Only after keys
are activated the secure channel is established, and sequence numbers are used
for MAC computation and verification. For each BPP packet, the sequence numbers
in bold must match at both peers; otherwise, the BPP packet is rejected.

\paragraph{SSH Binary Packet Protocol}
\label{subsec:ssh-bpp}

\begin{figure*}[tb]
    \centering
    \begin{subfigure}[t]{.99\columnwidth}
        \frame{\includegraphics[width=\linewidth]{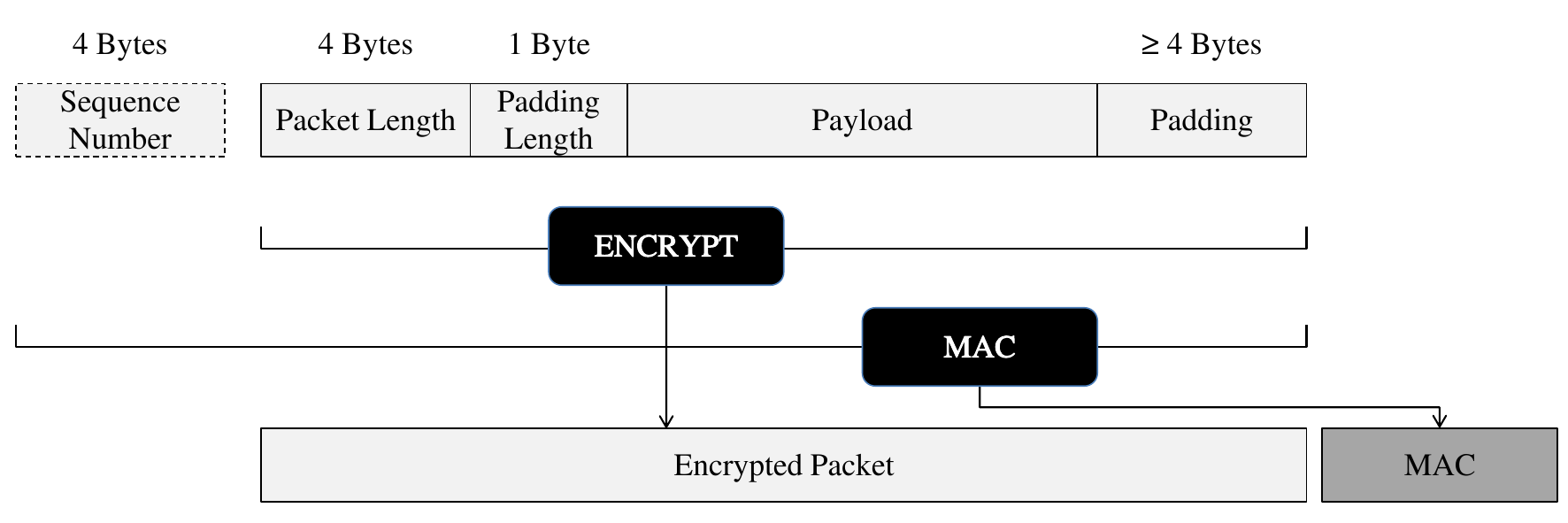}}
        \caption{Encrypt-and-MAC \cite{RFC4253} (diagram based on
        \cite[Fig.~1]{SP:AlbPatWat09}). The packet length is encrypted as part
        of the packet's plaintext. The sequence number is part of the input
        used to compute the MAC.}
        \label{fig:ssh-bpp-eam}
    \end{subfigure}
    \hspace{\columnsep}
    \begin{subfigure}[t]{.99\columnwidth}
        \frame{\includegraphics[width=\linewidth]{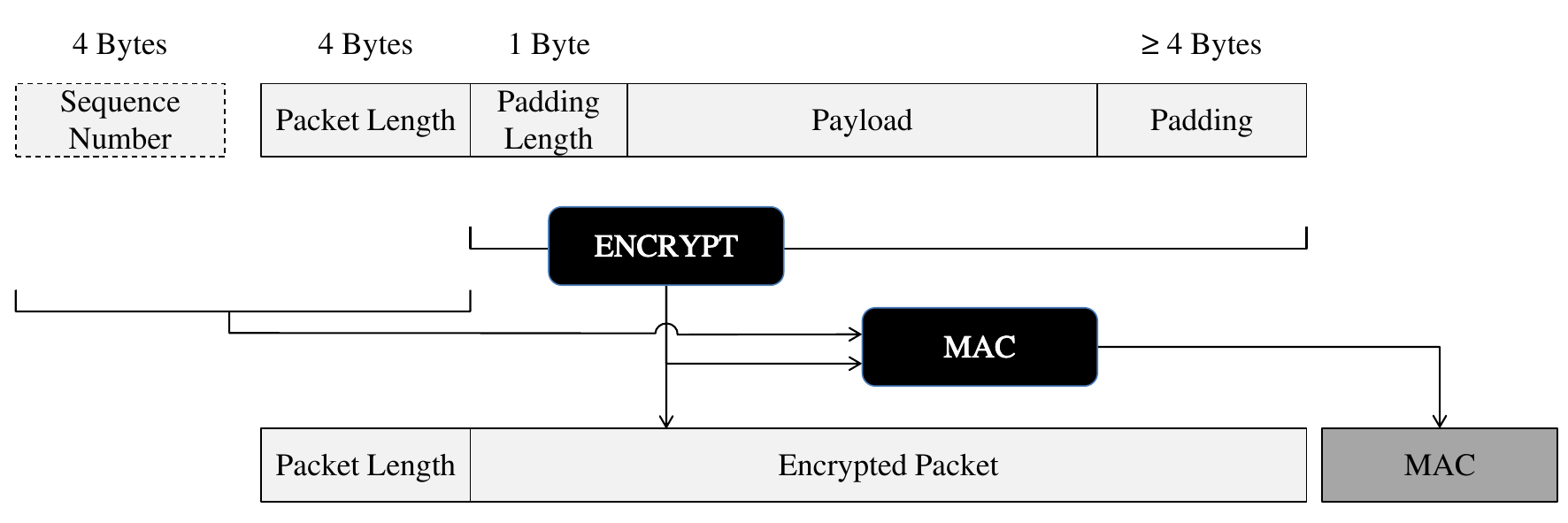}}
        \caption{Encrypt-then-MAC \cite[Sec.~1.5]{protocolopenssh} (diagram
        based on \cite[Slide~37]{Paterson2017}). The packet length is not
        encrypted but authenticated as part of the MAC's input. The sequence
        number is used similarly to Encrypt-and-MAC.}
        \label{fig:ssh-bpp-etm}
    \end{subfigure}
    \par\smallskip
    \begin{subfigure}[t]{.99\columnwidth}
        \frame{\includegraphics[width=\linewidth]{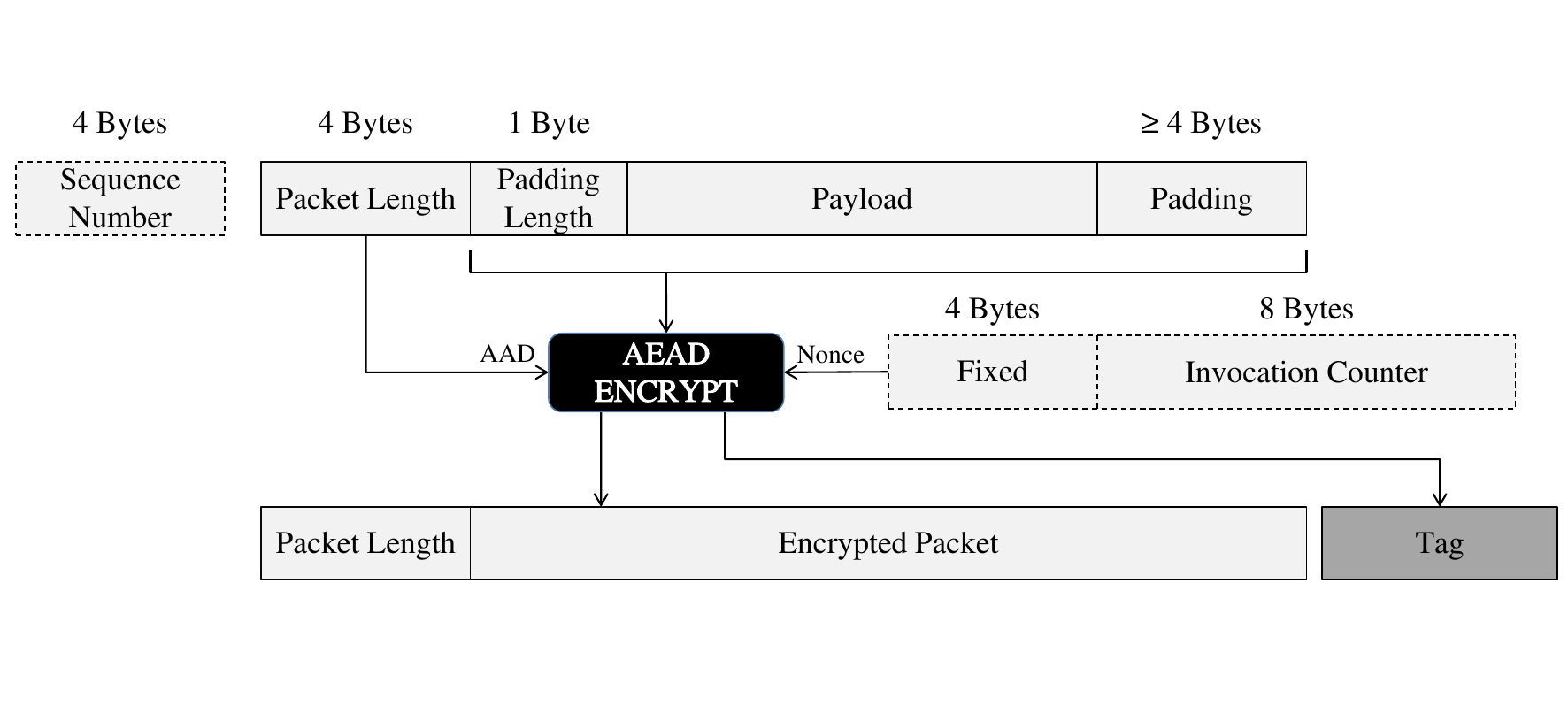}}
        \caption{Galois/Counter Mode \cite{RFC5647}. The packet length is not
        encrypted but authenticated as additional authenticated data (AAD). The
        sequence number is not used directly but is replaced with an invocation
        counter of constant offset.}
        \label{fig:ssh-bpp-gcm}
    \end{subfigure}
    \hspace{\columnsep}
    \begin{subfigure}[t]{.99\columnwidth}
        \frame{\includegraphics[width=\linewidth]{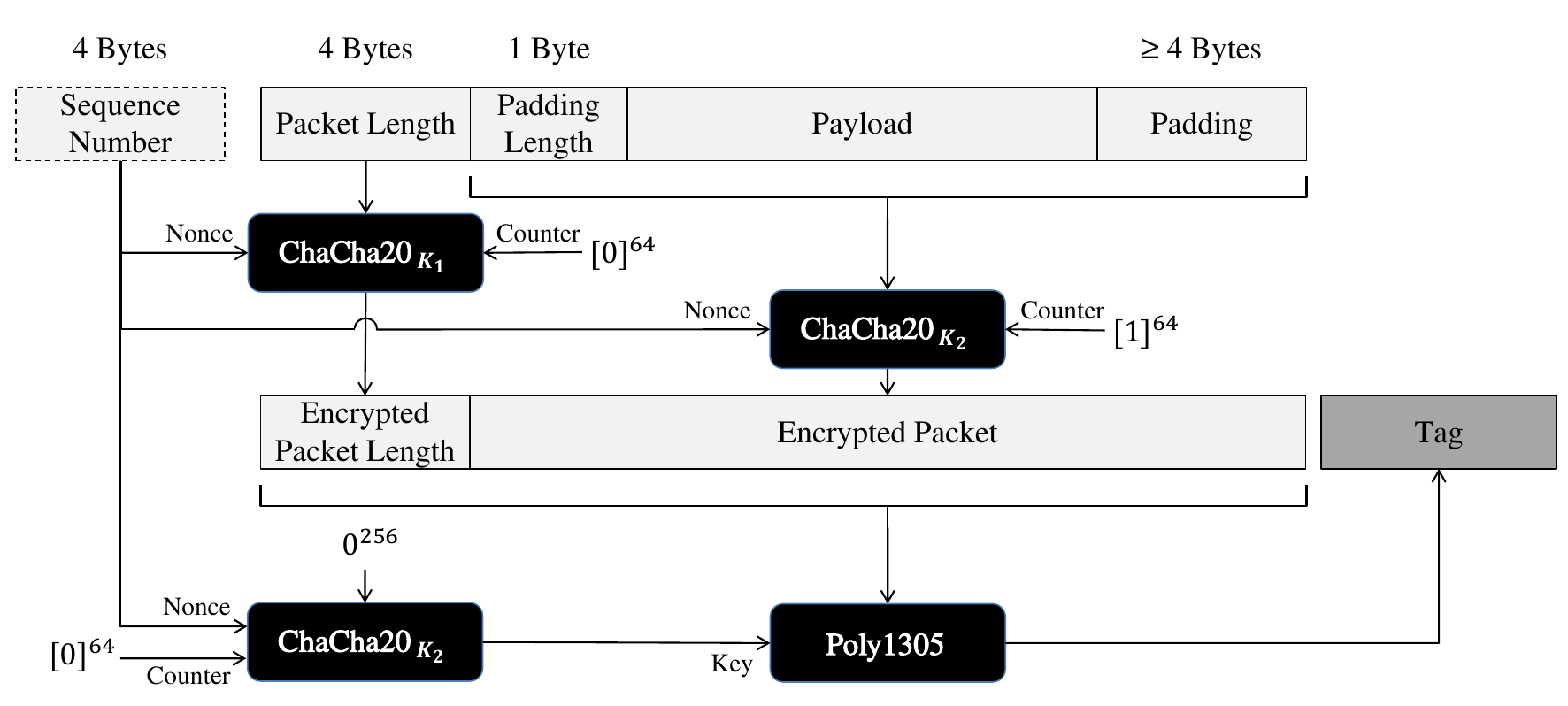}}
        \caption{ChaCha20-Poly1305 \cite{chachassh} (diagram based on
        \cite[Slide~39]{Paterson2017}). The packet length is encrypted using a
        different key and authenticated as part of the Poly1305 input. The
        sequence number is encoded as an unsigned 64-bit integer to match the
        nonce length of ChaCha20.}
        \label{fig:ssh-bpp-chacha20}
    \end{subfigure}
    \caption{Commonly used authenticated encryption schemes in the BPP of SSH.}
    \label{fig:ssh-bpp}
\end{figure*}

The BPP is used to encrypt and authenticate messages. First, a message is
prefixed by a 4-byte message length and a 1-byte padding length. Then, at least
4~bytes of padding are added to the message so that the total length is a
multiple of the block size or~8, whatever is larger. On the secure channel, the
packet is encrypted by the cipher mode, and a MAC is added. The details depend
on the authenticated encryption scheme, which uses an implicit initialization
vector \ivkdf derived from the session key.

CBC-EaM~\cite{RFC4253} (\autoref{fig:ssh-bpp-eam}) is part of the original SSH
specification. The MAC is computed over the implicit sequence number and the
packet plaintext. The IV of the first packet is \ivkdf, and IV chaining is used
(i.e., the IV of packet $i$ is the last ciphertext block of packet $i-1$).

CBC-EtM~\cite{protocolopenssh} (\autoref{fig:ssh-bpp-etm}) was added to OpenSSH
in 2012. Here, the packet length is \emph{not} encrypted to allow checking the
MAC before decryption. The MAC is computed over the sequence number, the
unencrypted packet length, and the ciphertext. The IVs are handled as with
CBC-EaM.

CTR~\cite{RFC4344} mode was proposed by Bellare, Kohno, and
Namprempre~\cite{CCS:BelKohNam02} as a countermeasure to attacks on CBC with IV
chaining. \ivkdf is used as the initial counter value and incremented after
encrypting a plaintext block. CTR can be used with EaM or EtM, with identical
implications for the length field and MAC computation as above.

GCM~\cite{RFC5647} (\autoref{fig:ssh-bpp-gcm}) mode was specified by the NSA
for Suite~B-compliant SSH implementations~\cite{RFC6239}. Here, ciphertext
integrity is part of the cipher mode. The length field is \emph{not} encrypted
(solely authenticated) to allow verification of the authentication tag before
returning any plaintext. Internally, GCM uses a 12-byte nonce that is
initialized to \ivkdf. The nonce is split into a 4-byte fixed value and an
8-byte invocation counter that is incremented by one for each message. The
sequence number is not used but is always offset by a constant from the
invocation counter.

ChaCha20-Poly1305~\cite{chachassh} (\autoref{fig:ssh-bpp-chacha20}) was added
to OpenSSH in 2013, inspired by a similar proposal for TLS by Langley and
Chang~\cite{agl-tls-chacha20poly1305-04,RFC7905}. Here, two different
encryption keys for the length field and the packet payload are derived, so the
length field cannot be used as a decryption oracle for the payload. The MAC is
computed over the concatenation of the two ciphertexts. Internally, the AEAD
construction uses the sequence number as a nonce for each packet.

We note that the SSH specification says that the length field is encrypted~\cite[Sec. 6]{RFC4253} and that the sequence number is used for integrity checks~\cite[Sec. 6.4]{RFC4253}. This is only true for CBC-EaM, CTR-EatM, and ChaCha20-Poly1305. The modes CBC-EtM, CTR-EtM, and GCM do not encrypt the length field, and GCM also does not 
use the sequence number.

%% file: sections/050-attacks.tex

In this section, we present a novel \emph{prefix truncation attack} on SSH. The
basic idea is that the attacker injects messages into the handshake to increase
the implicit sequence number in one of the peers and then deletes a
corresponding number of messages to that peer at the beginning of the secure
channel. Two key insights about the SSH protocol enable this attack:

\paragraph{SSH Does Not Protect the Full Handshake Transcript}
As detailed in~\autoref{sec:background}, the exchange hash signed by the server
during the handshake only authenticates some parts of the handshake transcript,
while other parts are left unauthenticated. This allows an attacker to inject
messages into the handshake, which cannot affect the key exchange but does
affect the implicit sequence numbers of the peers.

\paragraph{SSH Does Not Reset Sequence Numbers at the Beginning of the Secure Channel}
In SSH, sequence numbers are only incremented and never reset to~0, even when
the encryption key changes. This allows an attacker to manipulate the sequence
number counters in the secure channel before encryption and authentication keys
are activated.

\paragraph{Comparison to Other Protocols}
In IPsec/IKE, only a portion of the handshake transcript is signed, but unlike
SSH, sequence numbers are reset to~0 when encryption and MAC keys are
activated. In TLS, \textsc{Finished} messages are exchanged at the beginning of
the secure channel to verify the integrity of the complete handshake
transcript, and sequence numbers are reset to~0 after installing new keys. The
Noise Protocol Framework fully secures the handshake transcript and uses a
nonce as a sequence counter that is initialized to~0 after the handshake.

\subsection{Sequence Number Manipulation}
\label{sec:seqnomanipulation}

In this section, we first show how a \mitm attacker can arbitrarily increase
the receive sequence numbers \crcv and \srcv in the client and the server
during the handshake. This will be the basis for our prefix truncation attack
and its applications, allowing the attacker to compensate for messages deleted
from the secure channel.

\begin{figure}
    \centering
    \begin{subfigure}{.49\linewidth}
        \frame{\includegraphics[page=1,width=\linewidth]{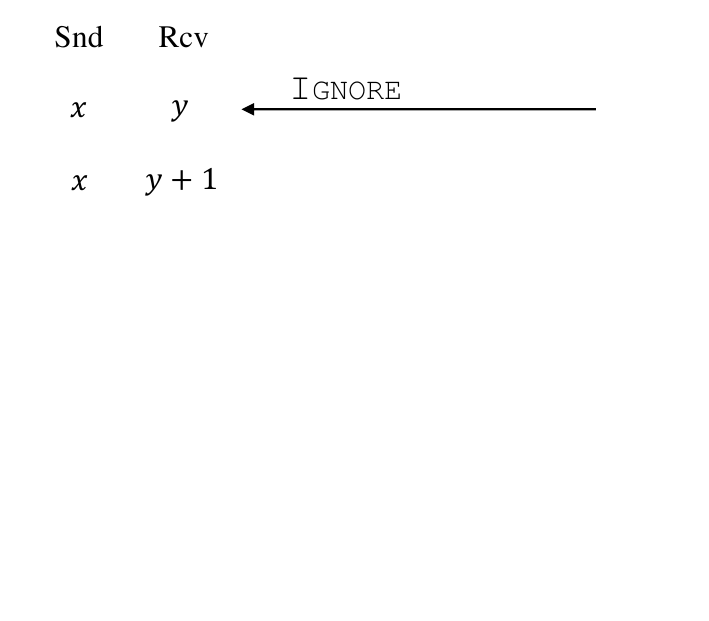}}
        \caption{\atkrcvinc}
        \label{fig:rcvincrement}
    \end{subfigure}
    \begin{subfigure}{.49\linewidth}
        \frame{\includegraphics[page=2,width=\linewidth]{img/Techniques.pdf}}
        \caption{\atkrcvdec}
        \label{fig:rcvdecrement}
    \end{subfigure}
    \begin{subfigure}{.49\linewidth}
        \frame{\includegraphics[page=3,width=\linewidth]{img/Techniques.pdf}}
        \caption{\atksndinc}
        \label{fig:sndincrement}
    \end{subfigure}
    \begin{subfigure}{.49\linewidth}
        \frame{\includegraphics[page=4,width=\linewidth]{img/Techniques.pdf}}
        \caption{\atksnddec}
        \label{fig:snddecrement}
    \end{subfigure}
    \caption{Techniques for sequence number manipulation as a \mitm in the SSH
    protocol. All techniques can target either client or server before the
    initial handshake concludes. The \mitm deletes all generated
    \msgunimplemented messages.}
\end{figure}

\paragraph{Technique \atkrcvinc (\autoref{fig:rcvincrement})}
A \mitm attacker can increase \crcv (resp. \srcv) by $N$ while not changing any
other sequence number by sending $N$ \msgignore messages to the client (resp.
server).

The correctness is evident from the fact that the SSH standard
requires for \msgignore that \emph{``All implementations MUST understand (and
ignore) this message at any time.''}~\cite[Sec. 11.2]{RFC4253}. The intended
purpose of this message is to protect against traffic analysis, so it is
considered a security feature, although there is no benefit from it during the
handshake phase. We note that the attacker may also use any other message type
that does not generate a response.

\paragraph{Other Modifications of Sequence Numbers}
In addition, we found that an attacker can set the sequence numbers to
arbitrary values by using the rollover after $2^{32}$ messages during the
handshake. These advanced techniques require that the implementation allows
handshakes with many messages, a large amount of data, and a long operating
time. We also require a message that generates a response message but is
otherwise ignored. Conveniently, the SSH standard requires this for all
messages with unrecognized message IDs~\cite[Sec.~11.4]{RFC5243}. Let
\msgunknown be a message with an unrecognized message ID.

\label{sec:advancedsqnnomanipulation}

\paragraph{Technique \atkrcvdec (\autoref{fig:rcvdecrement})}
A \mitm attacker can decrease \crcv (resp. \srcv) by~$N$ while not changing any
other sequence number by sending $2^{32}-N$ \msgignore messages to the client
(resp. server).

A single \msgignore message is only 5 bytes, so it fits into a single block
even for a 128-bit block cipher. Sending $2^{32}-N$ such messages transfers
$\approx 2^{32}\cdot 16\mathrm{B} \approx 69\mathrm{GB}$ of data. Consequently,
this technique can fail on implementations with timeouts or restrictions to the
amount of data or the number of messages transferred during the handshake.

\paragraph{Technique \atksndinc (\autoref{fig:sndincrement}) and \atksnddec (\autoref{fig:snddecrement})}
A \mitm can increase \csnd (resp. \ssnd) by $N$ while not changing any other
sequence number by sending $N$ \msgunknown and $2^{32} - N$ \msgignore messages
to the client (resp. server) and deleting all generated \msgunimplemented
messages. Conversely, a \mitm can decrease \csnd (resp. \ssnd) by $N$ while not
changing any other sequence number by sending $2^{32}-N$ \msgunknown and $N$
\msgignore messages to the client (resp. server) and deleting all generated
\msgunimplemented messages.

Here, the total data transfer required is $\approx 69\mathrm{GB}$ for
\atksndinc and twice as much ($\approx 138\mathrm{GB}$) for \atksnddec. Again,
these techniques may fail on implementations that have timeouts or restrict the
amount of data or number of messages exchanged during the handshake.

\paragraph{Evaluation}
We verified all techniques successfully against PuTTY~0.79. Additionally, our
experiments show that OpenSSH~9.5p1 recognizes a rollover of sequence numbers
and terminates the connection, thus not affected by any technique but
\atkrcvinc. AsyncSSH~2.13.2 and libssh~0.10.5 allow for \atkrcvinc but
terminate the connection due to handshake timeouts before any advanced
technique concludes. Dropbear~2022.83 disconnects on \msgunknown messages
instead of responding with \msgunimplemented but allows \xrcv to roll over,
therefore being affected by \atkrcvinc and \atkrcvdec only.

\subsection{Prefix Truncation Attack on the BPP}
\label{sec:ssh-attack}

\paragraph{Single Message Prefix Truncation Attack}
We assume the attacker wants to delete the first message \textsf{SC1} sent from
the server (\autoref{fig:ssh-attack}). The attack takes two steps:

\begin{enumerate}
    \item The attacker uses the \atkrcvinc technique to increase \crcv by one,
    e.g., by injecting an \msgignore message to the client
    before \msgnewkeys.

    \item The attacker deletes the first message \textsf{SC1} sent by the
    server.
\end{enumerate}

We first analyze this attack with regard to handshake authentication and
sequence numbers. As the key exchange does not protect the handshake transcript
from inserting \msgignore messages (\autoref{sec:background}), handshake
authentication is not broken. Before the first step, we have $\crcv = \ssnd$.
After the first step, we have $\crcv = \ssnd + 1$, but this manipulation is not
detected during the handshake. After the second step, we have $\crcv = \ssnd$,
and sequence numbers are back in sync.

It remains to be shown that the attacker can delete the message from the
channel, which requires knowledge about the message's length, and that its
deletion does not affect the MAC verification and decryption output for the
following messages. Both aspects require careful analysis with respect to the
used encryption mode, which will be given
in~\autoref{sec:ssh-attack:par:bytelength} and~\autoref{sec:vulnciphers}. Here,
we conclude by describing a straightforward generalization of the single
message attack.

\paragraph{$\mathbf{(N_S, N_C)}$-Prefix Truncation Attack}
In a single attack, the attacker can generally delete an arbitrary number of
$N_S$ initial messages sent from the server and $N_C$ initial messages sent
from the client. This is straightforward: Instead of inserting one \msgignore
message to the client before \msgnewkeys, the attacker inserts $N_S$ such
messages to the client and $N_C$ to the server. Consequently, instead of
deleting the first message from the server, the attacker deletes $N_S$ initial
messages from the server and $N_C$ initial messages from the client.

Note that the single message attack above is the specific case of a 
$(1,0)$-prefix truncation attack.

\subsection{Determining the Byte-Length of Messages}
\label{sec:ssh-attack:par:bytelength}
To successfully delete packets from the secure channel, the attacker has to
know their length. This is inherently true for encryption modes that do not
encrypt the packet length field (any EtM mode, GCM). In the case of an
encryption mode with an encrypted packet length field (any EaM mode,
ChaCha20-Poly1305), the attacker may employ different strategies to determine
the packet's length. One such strategy is to utilize knowledge about the
plaintext if the length of the first few messages inside the secure channel is
either fixed (for example, \msgserviceaccept) or can be measured within a
single connection ahead of time (for example, \msgextinfo). This approach was
used for all attacks described here. More advanced strategies may exploit TCP
segment sizes and timings, as well as the message order of the SSH protocol.
For example, an attacker may delay all encrypted traffic by the server until
after the client's \msgservicerequest message has been processed to determine
the length of the \msgextinfo message. Here, we assume that the attacker always
knows the lengths.

\subsection{Analysis of Encryption Modes}
\label{sec:vulnciphers}

\newcommand{\affectedTrue}{\ding{51}}
\newcommand{\affectedFalse}{\ding{55}}
\newcommand{\exploitableFalse}{\Circle}
\newcommand{\exploitableLimited}{\LEFTcircle}
\newcommand{\exploitableFull}{\CIRCLE}

\newcommand{\malleableSnd}{\ensuremath{\textcolor{purple}{\boldsymbol{\xsnd}}}}
\newcommand{\malleableRcv}{\ensuremath{\textcolor{purple}{\boldsymbol{\xrcv}}}}
\begin{table*}[tb]
    \centering
    \begin{tabular}{cccccccc}
        \toprule
        \multicolumn{2}{c}{Authenticated Encryption Mode} & Specification & Enc. State & Dec. State & Affected & Exploitable & Ref. \\
        \midrule
        \multirow{2}{*}{Encrypt-and-MAC} & CBC & \cite{RFC4253} & ($IV$, \malleableSnd) & ($IV$, \malleableRcv) & \affectedFalse & \exploitableFalse & \autoref{sec:vulnciphers:para:eam} \\
        ~ & CTR & \cite{RFC4253,RFC4344} & ($ctr$, \malleableSnd) & ($ctr$, \malleableRcv) & \affectedFalse & \exploitableFalse & \autoref{sec:vulnciphers:para:eam} \\\addlinespace
        \multirow{2}{*}{Encrypt-then-MAC} & CBC & \cite{RFC4253,protocolopenssh} & ($IV$, \malleableSnd) & ($IV$, \malleableRcv) & \affectedTrue & \exploitableLimited & \autoref{sec:vulnciphers:para:cbcetm} \\
        ~ & CTR & \cite{RFC4344,protocolopenssh} & ($ctr$, \malleableSnd) & ($ctr$, \malleableRcv) & \affectedTrue & \exploitableLimited & \autoref{sec:vulnciphers:para:ctretm} \\\addlinespace
        GCM & ~ & \cite{RFC5647} & $ctr_{Invocation}$ & $ctr_{Invocation}$ & \affectedFalse & \exploitableFalse & \autoref{sec:vulnciphers:para:gcm} \\\addlinespace
        ChaCha20-Poly1305 & ~ & \cite{chachassh} & \malleableSnd & \malleableRcv & \affectedTrue & \exploitableFull & \autoref{sec:vulnciphers:para:chacha20poly1305} \\
        \bottomrule
    \end{tabular}
    \caption{Authenticated encryption modes, corresponding specification
    documents, and their exposure to prefix truncation in the BPP of SSH. The
    initial value of state variables printed in bold purple can be chosen by
    the attacker, cf. \autoref{sec:seqnomanipulation}. Full control of either
    state enables perfect prefix truncation (\exploitableFull,
    ChaCha20-Poly1305). Partial control  may lead to limited exploitability,
    depending on the inner workings of the authenticated encryption mode
    (\exploitableLimited, Encrypt-then-MAC).}
    \label{tab:encryptionmodes}
\end{table*}

In this section, we analyze which encryption modes our attacks affect and if
they can be exploited in a real-world scenario
(see~\autoref{tab:encryptionmodes}). An encryption mode is \emph{affected} if,
after prefix truncation, all following packets on the secure channel are
decrypted, i.e., an AEAD mode does not generate the distinguished symbol
\textsc{Invalid} or a composed mode successfully verifies the MAC. Note that we
allow decryption to a different plaintext for probabilistic attacks. To capture
this, we define an encryption mode as \emph{exploitable} for an attack if the
message stream after decryption is well-formed and supports that attack. If the
attack's success probability is less than~1, we say the attack has
\emph{limited exploitability}.

\subsubsection{Not Affected}

\paragraph{GCM}
\label{sec:vulnciphers:para:gcm}
GCM~\cite{RFC5647} mode does not use the implicit sequence number. Instead, it
uses an invocation counter, initialized to \ivkdf, and incremented after each
message. The authors justify this by stating that the resulting nonce is always
a fixed offset from the sequence number. By deviating from the SSH standard,
GCM stops our attack, as the attacker cannot manipulate the invocation counter
during the handshake.

\paragraph{CBC-EaM and CTR-EaM}
\label{sec:vulnciphers:para:eam}
CBC uses IV chaining, and CTR uses a key stream. When the attacker deletes any
prefix of the ciphertext in either mode, the first ciphertext block received
will be decrypted as pseudorandom. Because EaM computes the MAC over the
plaintext, MAC verification will fail with a probability close to~1, thwarting
our attack.

\subsubsection{Affected And Perfectly Exploitable}

\paragraph{ChaCha20-Poly1305}
\label{sec:vulnciphers:para:chacha20poly1305}
ChaCha20-Poly1305~\cite{chachassh} directly uses the sequence number in its
internal key stream derivation, which makes it vulnerable to our prefix
truncation attack. All messages following the truncated prefix are decrypted to
their original plaintext because the integrity check of the AEAD cipher is done
over the ciphertext and the sequence number, which the attacker has manipulated
to match. Under the assumption that the attacker can correctly guess the packet
length, the prefix truncation attack always succeeds.

Note that the fault is not with ChaCha20-Poly1305 as an AEAD encryption scheme
but with its integration into the SSH secure channel construction.

\subsubsection{Affected With Limited Exploitability}

\paragraph{CTR-EtM}
\label{sec:vulnciphers:para:ctretm}
With CTR-EtM, the MAC is computed over the unencrypted length, the sequence
number, and the ciphertext. So, removing some packets from the beginning of the
channel does not cause a MAC failure, and cryptographically, the attack
succeeds. However, CTR uses a block counter initialized to \ivkdf, which
increments after each block. After prefix truncation, the key stream is
desynchronized, so \emph{all} following ciphertexts are decrypted as
pseudorandom packets. Each corrupted packet has a significant probability of
causing a critical failure, eventually stopping our attack.

\paragraph{Remark: Decryption Oracle for CTR-EtM Using Prefix Truncation}
For CTR-EtM, prefix truncation of~$k$ blocks (which exactly contain one or more
messages) provides a very limited \emph{decryption oracle} on the ciphertext
$c_1, \ldots, c_k$ where $c_i := \mathrm{Enc(\ivkdf + i)} \oplus p_i, 1\leq i
\leq k$. After deleting the first~$k$ blocks, MAC verification for the
following message of length~$l$ blocks will succeed because the length,
sequence number, and ciphertext are correct. The blocks $c_{k+1}, \ldots, c_{k
+l}$ will be decrypted as $p'_{j} := \mathrm{Enc(\ivkdf+j)} \oplus c_{k+j},
1\leq j \leq l$, and processed as a pseudorandom SSH message \textsf{SC1'}. Due
to format oracle side channels in SSH at the BPP layer, e.g., the padding
length, but also at the protocol layer, e.g., if a message is ignored or
triggers a response, the attacker can get some information about the bits
in~$p'_{j}$. This reveals information about the first~$l$ key stream blocks,
and thus also about $p_1, \ldots, p_l$, potentially leaking confidential
information like passwords in user authentication. If processing \textsf{SC1'}
does not cause a critical failure, the attack can even continue, revealing more
about the following key stream and, thus, plaintext. Exploiting this requires a
careful study of format oracles in SSH, which is outside the scope of this work.

\paragraph{CBC-EtM} \label{sec:vulnciphers:para:cbcetm}
With CBC-EtM, the MAC is computed from the unencrypted length, the sequence
number, and the ciphertext. The IV is not required because \ivkdf is implicit,
and all other IVs are authenticated before use. Consequently, prefix truncation
does not cause a MAC failure, and cryptographically, the attack succeeds.
Nevertheless, we need to consider the impact that IV chaining has on the
immediately following packet to see if this attack is practically exploitable.

Recall that the decryption of the first block is $p_1 := \mathrm{Dec}(c_1)
\oplus \ivkdf$, and for block $i$, it is $p_i := \mathrm{Dec}(c_i) \oplus  c_
{i-1}$. We assume the attacker uses prefix truncation to remove blocks $c_1,
\ldots, c_{k}$. The following block $c_{k+1}$ will now be decrypted as $p'_1 :=
\mathrm{Dec}(c_{k+1}) \oplus \ivkdf$. We are interested in how SSH
implementations process the resulting pseudorandom block $p'_1$ as the first
block in the decrypted packet. Intuitively, it should result in a corrupted
packet that causes a critical failure.\footnote{ Similarly to CTR-EtM, any
format oracle side channel for $p'_1$ reveals a relationship between \ivkdf and
$p_{k+1}$ via $\ivkdf \oplus p_{k+1} = c_{k} \oplus p'_1$, which is a marginal
information leak for the (secret) IV given information on $p_{k+1}$, and vice
versa. Again, we do not explore this further here.}

Surprisingly, there is a significant probability that the attack can continue,
although it is highly implementation-dependent. For a corrupted packet, there
are four possible outcomes:

\begin{enumerate}
    \item \emph{Critically Corrupt:} If corruption is detected at the BPP or
    application level, e.g., if a length field exceeds the packet length, the
    connection should be closed.

    \item \emph{Marginally Corrupt:} If the packet happens to be similar enough
    to the original, e.g., if the corruption is limited to optional fields, it
    should be processed without error and have the same effect as the original
    would have had.

    \item \emph{Evasively Corrupt:} If the packet is well-formed (i.e., has
    valid padding length) but has an unrecognized message~ID, an
    \msgunimplemented response must be sent, and the connection continues
    normally~\cite[Sec.~11.4]{RFC4253}.

    \item Any other case not covered above, in particular, recognized messages
    different from the original.
\end{enumerate}
Clearly, the first outcome stops any attack from going forward. However, the
second, third, and fourth outcomes may be beneficial for the attacker. We will
now present two instructive scenarios for outcomes two and three, and estimate
the success probability of an attack relying on that outcome. Later, we will
verify these estimates experimentally.

\paragraph{Scenario 1: CBC-EtM Prefix Truncation Of a Single Message, Second Message Has Format Flexibility}
In this scenario, the attacker wants to remove the first message, and the
second (corrupted) message needs to be functionally preserved but has some
format flexibility. For example, the second message might be \msgserviceaccept
(see~\autoref{sec:extdowngradeattack}), which is mandatory to start user
authentication. The encrypted part of the packet looks like this, where $p$ is
the padding length, $m$ is the message ID, and $n$ is the service name length:
\newlength{\maxheight}
\setlength{\maxheight}{\heightof{W}}
\newcommand{\baselinealignx}[1]{
         \centering
         \raisebox{0pt}[\maxheight][0pt]{#1}}
\begin{bytefield}[boxformatting=\baselinealignx,bitwidth=\linewidth/16]{16}
  \bitbox[]{1}{$p$}
  \bitbox[]{1}{$m$}
  \bitbox[]{4}{$n$}
  \bitbox[]{10}{Service Name}
  \\
  \bitbox{1}{\texttt{0e}} & \bitbox{1}{\texttt{06}} & \bitbox{4}{\texttt{00 00 00 0c}}
  \bitboxes*{1}{{\texttt{s}} {\texttt{s}} {\texttt{h}} {\texttt{-}}
  {\texttt{u}} {\texttt{s}} {\texttt{e}} {\texttt{r}} {\texttt{a}} {\texttt{u}}}
  \\
  \bitboxes*{1}{{\texttt{t}} {\texttt{h}}} \bitbox{14}{Random Padding}
\end{bytefield}

\noindent The probability that the first block decrypts exactly as shown is
only $2^{-128}$ for a 128-bit block cipher. However, for some clients, the
service name string is optional. These clients accept a 1-byte message with
$p=30$ ($\text{\texttt{0x1E}}$) and $m=6$ as \emph{marginally corrupt}, which
has a success probability of $2^{-16}$, independent of the block size.

Although \msgserviceaccept may be a lucky case (for the attacker), there are
structural reasons for this result: First, SSH messages are often short and can
be smaller than a single block. Second, the padding is random and cannot be
verified. Third, some messages have redundant fields that implementations
ignore (e.g., the service name above).

We experimentally verified that OpenSSH, Dropbear, PuTTY, and libssh allow
empty \msgserviceaccept messages from the server, enabling this attack. At the
same time, AsyncSSH is strict by requiring the correct service name.

\paragraph{Scenario 2: CBC-EtM Prefix Truncation Attack On More Than One Message}
In this scenario, we assume the attacker wants to remove the first $N>1$
messages and preserve all the following messages perfectly. Then, the attacker
can use prefix truncation to delete the first $N-1$ messages and take a bet on
the $N$-th message to be \emph{evasively corrupt}.

Let $\ell$ be the length of the ciphertext of the $N$-th message, with
padding length $p$, message ID $m$, and random padding. The attack succeeds
regardless of the content of the corrupted packet as long as it is well-formed
and unrecognized: A packet is \emph{well-formed} if $4 \leq p \leq \ell - 2$
(accounting for the padding length and message ID). A packet is \emph
{unrecognized} if $m$ is a message ID not known by the implementation.

Because the message is well-formed, it is not rejected at the BPP layer.
Furthermore, because the message is unrecognized, the peer must respond with
\msgunimplemented and otherwise ignore it~\cite[Sec.~11.4]{RFC4253}, so our
attack succeeds.

The probability that a packet is well-formed depends on~$\ell$. The padding
length is between~4 and 255, and $\ell$ is a multiple of $\mathrm{max}(8, \text
{block size})$, so the number of valid padding length values is $\mathrm{min}
(252, \ell - 5)$ out of $2^8$.

The probability that a packet is unrecognized depends on the size of the set
$U$ of unrecognized message IDs in the implementation. The attack requires at
least one unknown message ID. Through source code review, we identified~43~IDs
that are in active use, so we estimate up to 213 unknown message IDs out of
$2^8$.

In total, we estimate a success probability of $\mathrm{min}(252, \ell-5)\cdot
|U| \cdot 2^{-16}$. Assuming a block size of at least 128-bit (i.e., $\ell\geq
16$), we estimate that the success probability of this attack is between
$11\cdot 2^{-16} \approx 0.0002$ ($\ell_\mathrm{min}=16, |U_\mathrm{min}| = 1$)
and $252\cdot 213\cdot 2^{-16} \approx 0.8190$ ($\ell_\mathrm{max}\geq 252, |
U_\mathrm{max}| = 213$) for vulnerable implementations. Our experiments show
success probabilities from 0.0003--0.8383, in good agreement with our analysis
(\autoref{sec:extdowngradeattack}). Increasing the block size increases the
lower bound, while the upper bound stays the same.

%% file: sections/055-rw-attacks.tex

\label{sec:rwattack}

While the fact that BPP does not implement a secure channel is troublesome
enough, exploiting this vulnerability requires an analysis of the SSH protocol
after the handshake, i.e., the SSH authentication protocol.

As our attack achieves prefix truncation, it is natural to ask which SSH
messages can occur at the beginning of a secure channel. Historically, the
first messages exchanged are \msgservicerequest and \msgserviceaccept. Removing
either causes the connection to go stale, as the client will not begin the user
authentication. Then, our attack, while cryptographically successful, fails at
the application layer.

However, the SSH Extension Negotiation mechanism~\cite{RFC8308} introduces a
new message, \msgextinfo, which can occur immediately after \msgnewkeys as the
first message on the secure channel. Some of the extensions that can be
negotiated are security-relevant, providing an attack surface for our prefix
truncation attack and raising its impact.

In this section, we will first describe SSH Extension Negotiation and then
demonstrate how an attacker can downgrade the security of a connection by
removing the \msgextinfo message from the secure channel in a prefix truncation
attack.

\subsection{SSH Extension Negotiation}
\label{sec:extneg}
Even though the original SSH RFCs were designed with extensibility in mind,
they do not provide any mechanism to negotiate protocol extensions securely.
RFC 8308 \cite{RFC8308} closes this gap. The RFC describes a signaling
mechanism enabling extension negotiation, the extension negotiation mechanism
itself, and a set of initially defined extensions.

Support for extension negotiation is signaled as part of the \msgkexinit
message. The structure of the message is not altered, and the reserved field is
not used to avoid compatibility issues. Instead, each peer may include an
indicator name within the list of key exchange algorithms. The indicator name
differs depending on the role of the peer (\texttt{ext-info-c} vs.
\texttt{ext-info-s}) to avoid accidental negotiation.

Whenever a peer signals support for extension negotiation, the other side may
send an \msgextinfo message as the first message after \msgnewkeys.
Additionally, the server can send a second \msgextinfo later to authenticated
clients to avoid disclosing extension support to unauthenticated clients. Each
\msgextinfo message can contain several extension entries. Negotiation
requirements are defined on a per-extension level.

RFC 8308 defines an initial set of four protocol extensions, and vendors have
proposed and implemented additional extensions. We detail those relevant to our
attacks here.\footnote{We excluded the following extensions because we consider
them unrelated to our attacks: \texttt{no-flow-control},
\texttt{delay-compression}, \texttt{elevation}, \texttt{global-requests-ok},
\texttt{ext-auth-info}}

\texttt{server-sig-algs}~\cite{RFC8308} is a server-side extension that informs
the client about all supported signature algorithms when using a public key
during client authentication.

\texttt{publickey-hostbound@openssh.com}~\cite{protocolopenssh,agentrestrict}
is a server-side extension to advertise support for host-bound public key
authentication, which deviates from public key authentication by also covering
the server's host key. This allows the enforcement of per-key restrictions when
generating the signature outside the SSH client (i.e., when using SSH Agent).

\texttt{ping@openssh.com} \cite{protocolopenssh} is a server-side extension to
advertise support for a transport-level ping message similar to the Heartbeat
extension in TLS \cite{RFC6520}.

\subsection{Extension Downgrade Attack}
\label{sec:extdowngradeattack}
We now show how the prefix truncation attack can be applied to delete the
\msgextinfo message sent by the client, server, or both parties without either
noticing. Our attack differs depending on the encryption mode. For
ChaCha20-Poly1305, we can use the basic attack strategy. For CBC-EtM, we show
two strategies to generate additional messages in the secure channel so that
the attacker can use the ``evasively corrupt'' outcome of Scenario~2
in~\autoref{sec:vulnciphers:para:cbcetm}.

\paragraph{Impact}
Successfully performing the extension downgrade can directly impact the
security level of the connection. Most notably, the recently introduced
keystroke timing countermeasures by OpenSSH 9.5 will remain disabled when the
server has not sent \texttt{ping@openssh.com}. Furthermore, stripping an
\msgextinfo containing the \texttt{server-sig-algs} extension can lead to a
signature downgrade during client authentication, as the client has to resort
to trial-and-error instead.

\begin{figure*}[tb]
    \centering
    \begin{subfigure}[t]{.99\columnwidth}
        \frame{\includegraphics[page=1,width=\linewidth]{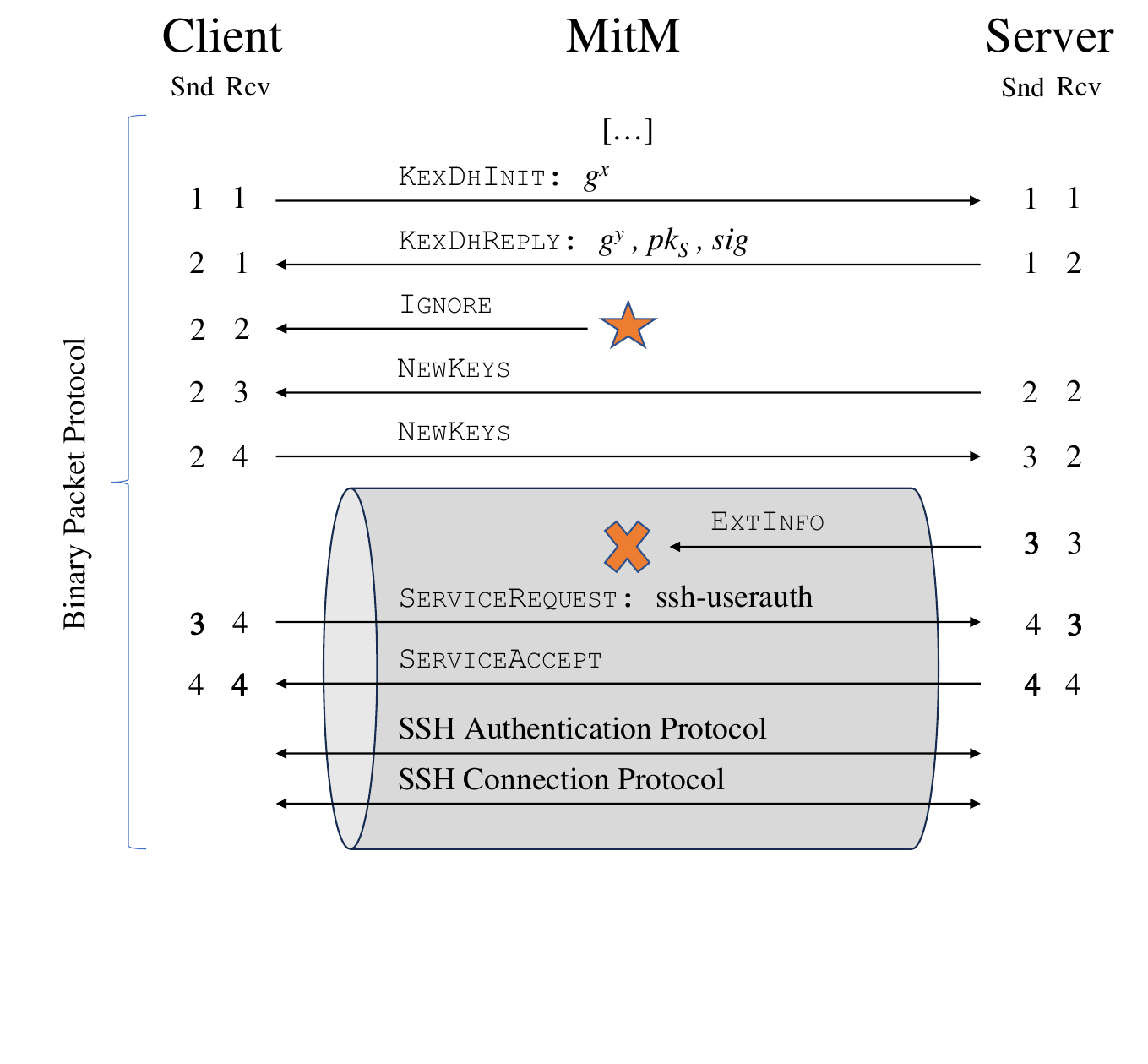}}
        \caption{Extension Downgrade Attack for ChaCha20-Poly1305: The \mitm
        injects an \msgignore message before the handshake concludes. The
        change in sequence numbers allows the \mitm to strip the \msgextinfo
        from within the secure channel.}
        \label{fig:extensiondowngradechacha}
    \end{subfigure}
    \hspace{\columnsep}
    \begin{subfigure}[t]{.99\columnwidth}
        \frame{\includegraphics[page=2,width=\linewidth]{img/Extension_Downgrade.pdf}}
        \caption{Extension Downgrade Attack for CBC-EtM: The \mitm injects
        \msgunknown before the \msgnewkeys is sent by the client. As the server
        already sent \msgnewkeys, the provoked \msgunimplemented message will
        be sent within the secure channel after \msgextinfo. The corrupted
        \msgunimplemented message has a significant probability of being
        ignored (see Scenario~2 in~\autoref{sec:vulnciphers:para:cbcetm}).}
        \label{fig:extensiondowngradecbc}
    \end{subfigure}
    \caption{Variants of the extension downgrade attack for ChaCha20-Poly1305
    and CBC-EtM.}
\end{figure*}

\paragraph{Extension Downgrade for ChaCha20-Poly1305}
The downgrade attack for ChaCha20-Poly1305 against the client is depicted in
\autoref{fig:extensiondowngradechacha}. It is identical to the single message
prefix truncation attack from~\autoref{sec:ssh-attack}, with \msgextinfo now
taking the place of \textsf{SC1} in~\autoref{fig:ssh-attack}. If the attack
should be directed against the server instead, a $(0,1)$-prefix truncation
attack should be performed. This allows an attacker to delete any \msgextinfo
sent immediately after \msgnewkeys.

While the server may send a second \msgextinfo just before signaling successful
client authentication, stripping the \msgextinfo message sent after \msgnewkeys
renders most publicly specified extensions unusable. This is because they are
either scoped to the authentication protocol, sent by the client only, or must
be sent by both parties to take effect. Solely the \texttt{ping@openssh.com}
extension may be sent in the second \msgextinfo to enable keystroke timing
countermeasures inside the connection protocol. However, OpenSSH~9.5 does not
implement any facility to send a second extension negotiation message. As shown
in \autoref{sec:internetscan}, extensions scoped to the authentication protocol
are the most common among SSH servers on the internet by a significant margin.

\paragraph{Extension Downgrade for CBC-EtM}
In~\autoref{fig:extensiondowngradecbc}, we show how the attack can also work
with CBC-EtM. Suppose an attacker injects an \msgunknown message to the server
after the server sends \msgnewkeys and \msgextinfo but before the client's
\msgnewkeys message (and also injects \msgunknown to the client to realign
sequence numbers). In that case, the server sends the response
\msgunimplemented as the second message in the secure channel immediately after
the \msgextinfo message. The attacker now wants to remove two messages from the
channel and can benefit from the ``evasively corrupt'' in Scenario~2
in~\autoref{sec:vulnciphers:para:cbcetm}. The attacker removes \msgextinfo from
the secure channel, which causes the decryption of the first block of
\msgunimplemented to become pseudorandom. Because \msgunimplemented messages
are relatively small ($\ell = 16$ for AES), the upper estimate for the success
probability is only $11 \cdot 213 \cdot 2^{-16} \approx 0.0358$.

However, the success probability can be increased significantly by exploiting
the new ping extension in OpenSSH~9.5. To make use of this, the attacker
replaces the \msgunknown message sent to the server with a \msgping message
containing at least 255 bytes of payload. As per specification, the server will
reflect this data in the \msgpong response. This yields $\ell \geq 264$, maxing
out the probability of the packet being well-formed. Consequently, the upper
estimate for the success probability is now $252 \cdot 213 \cdot 2^{-16}
\approx 0.8190$.

\paragraph{Evaluation}
We successfully evaluated the attack in 10,000 trials on ChaCha20-Poly1305 and
CBC-EtM against OpenSSH~9.5p1 and PuTTY~0.79 clients, connecting to
OpenSSH~9.4p1 (\msgunknown only) and 9.5p1. For CBC-EtM, our success rate in
practice was 0.0003 (OpenSSH) resp. 0.0300 (PuTTY), improved to 0.0074
(OpenSSH) resp. 0.8383 (PuTTY) when sending \msgping instead of \msgunknown.

%% file: sections/057-asyncssh.tex

\label{sec:asyncsshattacks}

Going beyond the SSH specifications, we now demonstrate how prefix truncation
attacks can also be used to exploit implementation flaws. Specifically, we
target AsyncSSH,\footnote{\url{https://github.com/ronf/asyncssh}} an SSH
implementation for Python with an estimated 60k daily
downloads.\footnote{\url{https://pypistats.org/packages/asyncssh}} We present
two attacks that exploit weaknesses in handling unauthenticated messages during
the handshake. These attacks are enabled by prefix truncation and sequence
number manipulation.

Note that we describe these attacks only for ChaCha20-Poly1305. Adjusting them
for CBC-EtM is straightforward, injecting appropriate \msgignore and
\msgunknown messages, but requires some of the advanced techniques described in
\autoref{sec:advancedsqnnomanipulation}. These advanced techniques only work
against some SSH implementations.

\subsection{Rogue Extension Negotiation Attack}
\label{sec:asyncrogueext}

\begin{figure}[tb]
    \centering
    \frame{\includegraphics[width=\columnwidth]{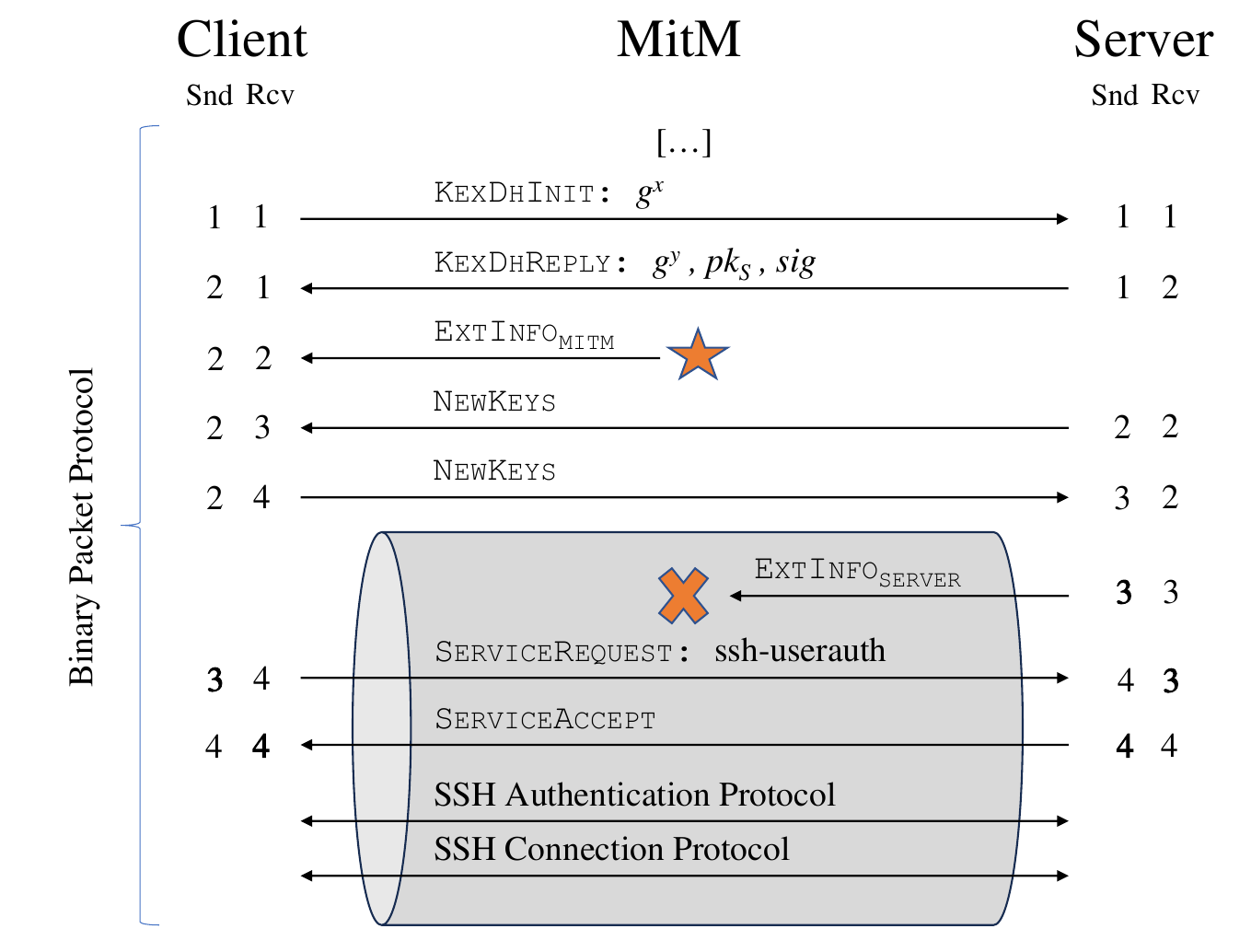}}
    \caption{Rogue Extension Negotiation Attack on AsyncSSH: The \mitm injects
    a malicious extension information message before the key exchange completes
    and deletes the server's \msgextinfo message to account for the change in
    sequence numbers. This attack relates to the generic extension downgrade
    attack in \autoref{sec:extdowngradeattack}.}
    \label{fig:rogueextnegattack}
\end{figure}

The rogue extension negotiation attack targets an AsyncSSH client connecting to
any SSH server sending an \msgextinfo message. The attack exploits an
implementation flaw in the AsyncSSH client to inject an \msgextinfo message
chosen by the attacker and a prefix truncation against the server to delete its
\msgextinfo message, effectively replacing it.

\paragraph{Impact}
The attacker can replace the content of the \msgextinfo message. AsyncSSH
clients support the \texttt{server-sig-algs} and \texttt{global-requests-ok}
extensions. Hence, the attacker can try to downgrade the algorithm used for
client authentication by restricting the value of \texttt{server-sig-algs} to a
subset of those supported by the server.

\paragraph{Attack Description}
The attack is a variant of the extension downgrade attack
in~\autoref{sec:extdowngradeattack}, but instead of \msgignore, the attacker
sends a chosen \msgextinfo packet to the client. Similar to \msgignore,
\msgextinfo does not trigger a response from the client. A correct SSH
implementation should not process an unauthenticated \msgextinfo message.
However, the injected message is accepted due to flaws in AsyncSSH.

\paragraph{Evaluation}
We successfully evaluated the attack against AsyncSSH~2.13.2 as a client,
connecting to AsyncSSH~2.13.2.

\subsection{Rogue Session Attack}
\label{sec:asyncrogueshell}

\begin{figure}[tb]
    \centering
    \frame{\includegraphics[width=\columnwidth]{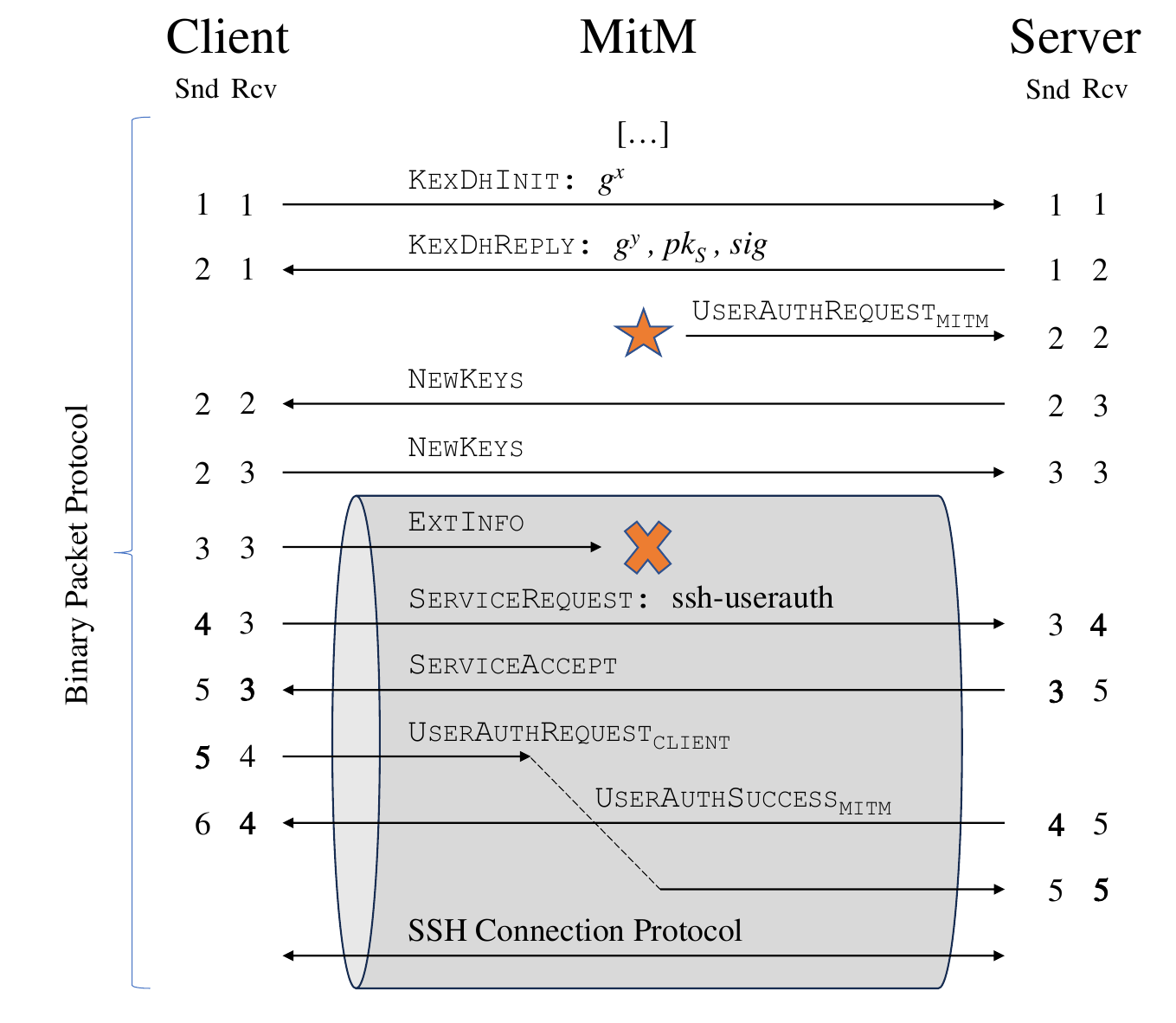}}
    \caption{Rogue Session Attack on AsyncSSH: The \mitm injects a malicious
    authentication request before the handshake is complete and deletes the
    client's \msgextinfo message to account for the change in sequence numbers.
    By delaying the authentication request sent by the client, the \mitm
    ensures that the malicious one is processed. Any additional authentication
    requests are silently ignored.}
    \label{fig:roguesessionattack}
\end{figure}

The rogue session attack targets any SSH client connecting to an AsyncSSH
server, on which the attacker must have a shell account. The attack's goal is
to log the client into the attacker's account without the client being able to
detect this.

\paragraph{Impact}
With a successful attack, the attacker can gain complete control over the
remote end of the SSH session. The attacker receives all keyboard input by the
user, completely controls the terminal output of the user's session, can send
and receive data to/from forwarded network ports, and can create signatures
with a forwarded SSH Agent, if any. The result is a complete break of the
confidentiality and integrity of the secure channel, providing a strong vector
for a targeted phishing campaign against the user. For example, the attacker
can display a password prompt and wait for the user to enter the password,
elevating the attacker's position to a \mitm at the application layer and
enabling impersonation attacks.

\paragraph{Attack Description}
The messages exchanged during the attack are depicted in
\autoref{fig:roguesessionattack}. The attacker injects a chosen
\msguserauthrequest before the client's \msgnewkeys. This request must be a
valid authentication request containing the credentials of the attacker. The
attacker can use any authentication mechanism that does not require exchanging
additional messages between client and server, such as \texttt{password} or
\texttt{publickey}. Due to a state machine flaw, the AsyncSSH server accepts
the unauthenticated  \msguserauthrequest message and defers it until the client
has requested the authentication protocol.

To avoid a race condition between the \msguserauthrequest sent by the client
and the \msguserauthrequest injected by the attacker, the attacker delays the
client's \msguserauthrequest until after the server signals a successful
authentication in response to the injected \msguserauthrequest. The AsyncSSH
server silently ignores any additional authentication request after a
successful authentication.

To complete the attack, the attacker has to fix the sequence numbers using one
of two strategies (note that \autoref{fig:roguesessionattack} only shows the
first strategy):

\begin{itemize}
    \item Suppose the client sends an extra message before \msgservicerequest.
    In that case, the attacker can delete that message from the channel,
    effectively performing the (0,1)-prefix truncation attack with
    \msguserauthrequest instead of the usual \msgignore message.

    \item Alternatively, suppose the server sends an extra message before
    \msgserviceaccept. In that case, the attacker can delete that message after
    injecting an additional \msgunknown message to the client before
    \msgnewkeys, triggering an \msgunimplemented response that is deleted. This
    increases both \csnd and \crcv, moving the send count deficit from the
    client to the server.
\end{itemize}

\paragraph{Evaluation}
We successfully evaluated the attack against AsyncSSH~2.13.2 as a server,
connecting to AsyncSSH~2.13.2 and OpenSSH~9.4p1.

%% file: sections/065-statistics.tex

\label{sec:internetscan}

To estimate the impact of the prefix truncation attacks, we scan for the SSH
servers preferring or supporting any affected encryption mode. Similarly, to
estimate the impact of the extension downgrade attack, we scan for servers
sending \msgextinfo messages.

\paragraph{Methodology}
For scanning, we used ZMap \cite{USENIX:DurWusHal13} and ZGrab2
\cite{CCS:DAMBH15} on port~22 of the entire IPv4 address space. The scan was
performed over two days in early October 2023, totaling 15.164M SSH servers.

As ZGrab2 cannot capture SSH extensions, we performed a complementary scan at
the end of June 2023, using a custom tool, on a subset of $2^{20}$ open ports.
The scan covered a total of 830k servers. All data relating to the use of
extension negotiation in SSH is sourced from this scan.

In SSH, the algorithm order of the client determines which algorithm is
preferred. However, we cannot scan for actual client use. Assuming that servers
and clients are bundled in a single product and share algorithm preference and
support, we use the server's lists as a surrogate, as was also done
in~\cite{CCS:ADHP16}.

\paragraph{Symmetric Encryption Algorithms}
In~\autoref{tab:sshciphersusage}, we show the number of servers that prefer and
support various encryption modes. A cipher is preferred if it is placed first
in the list of supported algorithms.

We find that, by far, the most preferred encryption cipher is
ChaCha20-Poly1305, with 57.64\% listing this algorithm first. This is followed
by AES-CTR (31.56\%) and, with some distance, by AES-GCM (8.04\%) and AES-CBC
(1.56\%).

\newcommand{\algname}[1]{#1}
\begin{table}[tb]
    \centering
    \setlength{\tabcolsep}{3.5pt}
    \small
    \begin{tabular}{lrrrr}
        \toprule
        Cipher Family & \multicolumn{2}{c}{Preferred} & \multicolumn{2}{c}{Supported} \\
        \midrule
        \algname{ChaCha20-Poly1305} & 8,739k & 57.64\% & 10,247k & 67.58\% \\
        \algname{AES-CTR} & 4,785k & 31.56\% & 14,866k & 98.04\% \\
        \algname{AES-GCM} & 1,219k & 8.04\% & 10,450k & 68.92\% \\
        \algname{AES-CBC} & 236k & 1.56\% & 4,069k & 26.84\% \\
        Other & 147k & 0.97\% & - & - \\
        Unknown / No \msgkexinit & 34k & 0.23\% & - & - \\
        \midrule
        Total & 15,164k & 100\% & & \\
        \bottomrule
    \end{tabular}
    \caption{Preferred SSH cipher families as of October 2023.}
    \label{tab:sshciphersusage}
\end{table}

\paragraph{Authenticated Encryption Modes}
As non-AEAD ciphers must be combined with a MAC, we also evaluate which
\emph{authenticated} encryption modes the servers prefer and support. The
numbers for the AEAD modes ChaCha20-Poly1305 (57.64\%) and GCM (8.04\%) are
identical to those for encryption modes, as the MAC is already integrated.
Preference for CTR modes is split between a majority for CTR-EaM (26.14\%) and
a minority for CTR-EtM (5.46\%). Preference for CBC modes is mostly CBC-EaM
(2.37\%), while preference for CBC-EtM (0.09\%) is marginal.

In summary, \scanPreferred\% of all servers prefer an authenticated encryption
mode affected by our attacks.

Looking at the support for authenticated encryption modes vulnerable to our
attacks, we find that 67.58\% of all servers support ChaCha20-Poly1305, while
17.24\% support CBC-EtM. In total, \scanSupported\% support at least one
affected mode.

\begin{table}[tb]
    \centering
    \setlength{\tabcolsep}{4pt}
    \small
    \begin{tabular}{lrrrr}
        \toprule
        AE Mode & \multicolumn{2}{c}{Preferred} & \multicolumn{2}{c}{Supported} \\
        \midrule
        \algname{ChaCha20-Poly1305} & 8,739k & 57.64\% & 10,247k & 67.58\% \\
        \algname{CTR-EaM} & 3,964k & 26.14\% & 4,200k & 27.70\% \\
        \algname{GCM} & 1,219k & 8.04\% & 10,450k & 68.92\% \\
        \algname{CTR-EtM} & 828k & 5.46\% & 10,685k & 70.46\% \\
        \algname{CBC-EaM} & 359k & 2.37\% & 1,585k & 10.46\% \\
        \algname{CBC-EtM} & 14k & 0.09\% & 2,614k & 17.24\% \\
        Other & 2k & 0.01\% & - & - \\
        Unknown / No \msgkexinit & 36k & 0.24\% & - & - \\
        \midrule
        Total & 15,164k & 100\% & & \\
        \bottomrule
    \end{tabular}
    \caption{Distribution of supported authenticated encryption modes as of
    October 2023.}
    \label{tab:sshciphermodes}
\end{table}

\paragraph{SSH Extensions}
We also looked at SSH extensions offered by servers before user authentication;
see~\autoref{tab:sshextensions}. We can see that 76.81\% of all servers send
the \texttt{server-sig-algs} extensions to indicate support for better
signature schemes for client public key authentication. Furthermore, 8.8\% send
the \texttt{publickey-hostbound} extension, improving security for
authentication using SSH agent. Both extensions provide opportunities for
downgrade attacks, as their absence can weaken the strength of the
authentication.

\begin{table}[tb]
    \centering
    \begin{tabular}{lrr}
         \toprule
         Extension name &  \multicolumn{2}{c}{Times Offered} \\
         \midrule
         \texttt{server-sig-algs} & 637,466 & 76.81\% \\
         \texttt{publickey-hostbound@} & 73,040 & 8.80\% \\
         \texttt{delay-compression} & 283 & 0.03\% \\
         \texttt{no-flow-control} & 283 & 0.03\% \\
         \texttt{global-requests-ok} & 283 & 0.03\% \\
         \bottomrule
    \end{tabular}
    \caption{SSH extensions offered by servers after the initial handshake,
    \texttt{@openssh.com} abbreviated to \texttt{@}. Extensions sent by servers
    upon successful client authentication are not included.}
    \label{tab:sshextensions}
\end{table}

%% file: sections/070-countermeasures.tex

\label{sec:countermeasures}

As a stop-gap measure, the affected cipher modes can be turned off. Widely
supported alternatives are AES-GCM or AES-CTR. However, the root cause analysis
shows that the underlying issues lie in the SSH specification. We therefore
suggest two changes to the specification.

\paragraph{Sequence Number Reset}
Resetting sequence numbers to zero when encryption keys are activated ensures
that sequence number manipulations during the handshake can no longer affect
the secure channel. Unfortunately, sequence number reset is a major break in
compatibility. To avoid connection failures due to one-sided sequence number
resets, we suggest that an implementation signals the support for this
countermeasure by including an identification string in the list of supported
key exchange algorithms. The SSH extension negotiation mechanism is already
employing this method. If and only if both peers signal support for this
countermeasure, the sequence numbers will be reset.

In response to our findings, OpenSSH implemented this behavior as part of their
so-called \enquote{strict kex} countermeasure
\cite[Sec. 1.10]{protocolopenssh}. In addition to resetting sequence numbers,
\enquote{strict kex} mandates that unexpected or unknown messages during the
initial key exchange must lead to the connection's termination. An unexpected
message in this context is any message that is not strictly required for key
exchange. \enquote{strict kex} has since been adopted by various vendors to
ensure interoperability between SSH implementations.

\paragraph{Full Transcript MAC}
Authenticating the full handshake transcript, as seen by the client and server,
can detect attempts of handshake manipulation by a \mitm attacker, including
sequence number manipulation through our techniques. It is impossible to extend
the scope of the existing exchange hash, as the server signature is transmitted
before the new keys are taken into use. Therefore, any messages sent after the
key exchange but before \msgnewkeys cannot be included. We suggest that both
peers send a MAC authenticating the entire transcript at the start of the
channel, similar to TLS Finished messages. Signaling support should be done as
above. However, the transcript must be carefully canonicalized. While client
and server messages are sequential, they can interleave asynchronously, leading
to transcript variations. Also, the protocol must be extended to define the
algorithm, encoding, and position of the transcript MAC. Thus, securing the
handshake is more complex than resetting the sequence number.

\newcommand{\msgcount}{\ensuremath{M}}
\newcommand{\binspace}{\ensuremath{\{0,1\}^*}}

\paragraph{Relationship to Formal Proofs}
Both countermeasures have a common goal: Align the SSH standard with
expectations for stateful encryption schemes from formal models for the BPP
presented in~\cite{CCS:BelKohNam02,CCS:ADHP16}. A sequence number reset
achieves this directly by initializing the sequence numbers to zero, as in the
models. On the other hand, verifying the full transcript hash forces the
sequence number in the stateful encryption and decryption methods to be
synchronized by the sender and receiver. Although the sequence numbers are then
not initialized to zero, each pair is nevertheless initialized to a common
value out of the attacker's control. The existing models could then be adjusted
in the following way: If $T_C, T_S \in \binspace$ are the (canonicalized)
transcripts of the SSH handshake as seen by the client and the server, and
$\msgcount_{CS},\msgcount_{SC}:\binspace\mapsto \mathbb{N}$ are functions
counting the messages from the client to the server and vice versa in a
transcript, then sequence numbers in the stateful encryption and decryption
modes are initialized to:
\begin{align*}
    \csnd &= \msgcount_{CS}(T_C), &
    \crcv &= \msgcount_{SC}(T_C),\\
    \ssnd &= \msgcount_{SC}(T_S), &
    \srcv &= \msgcount_{CS}(T_S).
\end{align*}
Authenticating the transcript then ensures that $T_C = T_S$, and thus
$\csnd=\srcv$ and $\crcv=\ssnd$, before the first messages in the secure
channel are encrypted or decrypted. Authenticating the handshake transcript has
the added benefit that the handshake could be analyzed in a \enquote{matching
conversations}-based security model~\cite{C:JKSS12,C:KraPatWee13}.

\paragraph{Other Issues}
We suggest that SSH specifies \enquote{end-of-communication} messages to detect
suffix truncation attacks. Also, AsyncSSH should be hardened to disallow
unauthenticated, application-layer messages during the SSH handshake. In
response to our findings, the state machine of AsyncSSH was improved in
version~2.14.1 to mitigate our attacks.

%% file: sections/075-future.tex

Formally, SSH BPP security was modeled as stateful decryption
\cite{CCS:BelKohNam02,EC:PatWat10,CCS:ADHP16}. Implicitly, this state was
associated with SSH sequence numbers, and it was assumed that an adversary
could not manipulate this state. These models can be extended in two
directions: (1) Include a broader definition of state. By including chained
IVs, key stream state, and GCM invocation counters, these models can be used to
show why certain cipher modes resist our attacks and that they indeed achieve
INT-PST security. (2) Introduce a novel adversarial query,
\texttt{ModifyState}, to model the attacks described here.

Our attack combines weaknesses in the SSH handshake with weaknesses in the
encrypted channel. Earlier work analyzed these separately, leading to small
models. To find our attack automatically, models of SSH for computer-aided
proofs could (1) model the handshake as well as the BPP together, (2) keep
track of sequence numbers in the BPP, including the handshake, which requires
modeling integer numbers that can overflow as the internal state, (3) model
seemingly unimportant messages like \msgignore, and (4) consider each
encryption mode separately. The properties to verify should include strong
security notions such as INT-aPTXT~\cite{C:FGMP15}.

Applying state learning to implementations also has the potential to find our
attacks automatically in the future, although it suffers from a combinatorial
explosion in the number of messages (see Section 5.1 in~\cite{lesiuta2022}).
Messages like \msgignore and \msgextinfo need to be included in the alphabet to
find our attacks, and an active \mitm attacker has to be considered.

%% file: sections/080-conclusion.tex

We have shown that the complexity of SSHv2 has increased over its 25~years of
development to a point where the addition of new algorithms and features has
introduced new vulnerabilities. The root cause analysis has shown that the
potential for our attacks was already present in the original specification.
Handshake transcripts were never fully authenticated, and sequence numbers were
never reset to~0. However, as new authenticated encryption modes and extension
messages were added, these weaknesses grew into exploitable vulnerabilities.

We introduced novel sequence number manipulation and prefix truncation attacks
for secure channels, which invalidate the INT-aPTXT~\cite{C:FGMP15} security of
SSH BPP for certain ciphers. We extended these vulnerabilities to real-world
exploits like disabling SSH extension negotiation. This yields novel insights
into the complex interplay between a practical security mechanism (sequence
numbers) and abstract security notions (INT-PTXT vs. INT-CTXT,
\cite{JC:BelNam08}).

Our close look at the extension negotiation mechanism reveals its design
weaknesses: First, sending \msgextinfo is optional even if both parties signal
support for extension negotiation during the handshake. Second, \msgextinfo
cannot be used to change the SSH handshake itself, e.g., to implement the
countermeasures proposed in this paper. However, it outperforms extension
negotiation within the \msgkexinit in aspects of privacy as protocol extension
can be negotiated securely, i.e., privately, similar to encrypted extensions in
TLS 1.3. As a consequence, extension negotiation within the \msgkexinit should
be strictly limited to extensions affecting the SSH handshake. Protocol
extensions affecting the user authentication or application layer should be
negotiated through the extension negotiation mechanism.

Although we suggest backward-compatible countermeasures to stop our attacks,
the security of the SSH protocol could benefit from a redesign from scratch.
The redesign process could be inspired by that of TLS~1.3, which brought
implementers together with experts in protocol analysis and formal
verification~\cite{SP:BBBBCLP21}. This could simplify the protocol while
preserving and/or achieving desired security notions for SSH, which may differ
from those of TLS. For example, while the privacy of client authentication and
extension negotiation are relatively new features for TLS, they are already
present in SSH and should thus be preserved in a redesign.

%% file: sections/090-ack.tex

Fabian Bäumer was supported by the German Federal Ministry for Economic Affairs
and Climate Action (BMWK) project \enquote{Industrie 4.0 Recht-Testbed}
(13I40V002C). Marcus Brinkmann was supported by the Deutsche
Forschungsgemeinschaft (DFG, German Research Foundation) under Germany's
Excellence Strategy - EXC 2092 CASA - 390781972.

%% file: ae/appendix.tex
\appendix
\section{Artifact Appendix}

\subsection{Abstract}

This document describes the artifacts to the USENIX Security '24 Publication
\emph{Terrapin Attack: Breaking SSH Channel Integrity By Sequence Number
Manipulation}.

Using these instructions, the evaluations of Sequence Number Manipulation
(Sect.~4.1), Extension Downgrade Attack (Sect.~5.2), Rogue Extension Attack
(Sect.~6.1) and Rogue Session Attack (Sect.~6.2) can be reproduced.

Also, the aggregation scripts for the internet scans are available and can be
tested on a small subset of the samples.

\subsection{Description \& Requirements}

\subsubsection{Security, privacy, and ethical concerns}
The configuration uses the host network to allow (optional) monitoring of the
attack using Wireshark or other network packet analysis tools on the loopback
interface. During the runtime of the evaluation, this makes the tested SSH
server and proof of concept (PoC) available to all systems with access to the
local network (TCP bind to \verb|0.0.0.0|, ports 2200 and 2201). Reviewers
should take care to isolate the test system from the internet, for example
using a firewall.

\subsubsection{How to access}
The artifacts are publically available at
\url{https://github.com/RUB-NDS/Terrapin-Artifacts/tree/9907c80fa7e4184a29ceac352947ea51a49dce6a}.

\subsubsection{Hardware dependencies}
None.

\subsubsection{Software dependencies}
\begin{itemize}
    \item Linux or MacOS.\footnote{Windows WSL might work but is untested and
    \emph{not} supported.} No specific distribution or version is required. We
    used Manjaro (rolling release in March 2024) and MacOS~14.4 (Sonoma).

    \item Bash shell interpreter (typically included in the above). No specific
    version is required. We used bash~5.2.26 and 3.2.57.

    \item Docker Engine or Docker Desktop. While Docker Engine suffices and is
    typically included in Linux distributions, Docker Desktop is a separate
    install on MacOS. No specific version is required. We used Docker
    Engine~25.0.3 and Docker Desktop~4.28.0.

    \item Wireshark (optional), for network packet analysis.
\end{itemize}

\subsubsection{Benchmarks}
None.

\subsection{Set-up}
All required Docker images are built on demand when the evaluation scripts are
executed, so no setup is required.

The TCP ports 2200 and 2201 should be free and available. This is the case by
default on many systems. Some systems might require a configuration of the
firewall to allow the test servers to bind \verb|0.0.0.0| on these ports. On
some systems, the firewall will show a pop-up dialog when the first server
starts up, requiring manual confirmation.

\subsubsection{Installation}

\paragraph{Linux:}
Install the Docker engine under a supported Linux distribution by following the
instructions available at \url{https://docs.docker.com/engine/install/}.

\paragraph{MacOS:}
Install Docker Desktop available at \url{https://www.docker.com/products/docker-desktop/}.

\subsubsection{Basic Test}

The following scripts build all required Docker images and can be used as a
basic functionality test. It will also be called by all evaluation scripts, so
this step is optional.

\begin{lstlisting}
$ impl/build.sh
[+] Building 2.0s (15/15) FINISHED
[...]
=> => naming to docker.io/terrapin-artifacts/openssh-server:9.4p1
[...]
$ pocs/build.sh
[...]
\end{lstlisting}

The output shows the progress on downloading base images and building the
evaluation images. If there is no output, all docker images are already built.

\subsection{Evaluation workflow}

\subsubsection{Major Claims}
We evaluated our attacks against several clients using an OpenSSH~9.5p1
(C1, C2) or AsyncSSH~2.13.2 (C3, C4) server. For an overview of the expected
outcomes, see also~\autoref{tbl:outcomes}.

\newcommand{\Rtit}[1]{\rlap{\rotatebox{45}{#1}}}

\newcommand{\aNot}{-}
\newcommand{\aSuc}{\checkmark}
\newcommand{\aKlR}{R}
\newcommand{\aKlT}{T}
\newcommand{\aKlU}{U}
\newcommand{\aUnk}{\,\,\,\,\,}

\begin{table}
    \setlength{\tabcolsep}{0.3em}
    \small
    \centering
    \begin{threeparttable}
        \begin{tabularx}{\linewidth}{lcccccc}
            \toprule
            \textbf{Attack}
            & \Rtit{PuTTY~0.79}
            & \Rtit{OpenSSH~9.4p1}
            & \Rtit{OpenSSH~9.5p1}
            & \Rtit{Dropbear~2022.83}
            & \Rtit{AsyncSSH~2.13.2}
            & \Rtit{libssh~0.10.5}
            \\
            \midrule
            C1 \atkrcvinc & \aSuc & \aNot & \aSuc & \aSuc & \aSuc & \aSuc \\
            C1 \atkrcvdec & \aSuc & \aNot & \aKlR & \aSuc & \aKlT & \aKlT \\
            C1 \atksndinc & \aSuc & \aNot & \aKlR & \aKlU & \aKlT & \aKlT \\
            C1 \atksnddec & \aSuc & \aNot & \aKlR & \aKlU & \aKlT & \aKlT
            \vspace{0.3em}
            \\
            C2 ChaCha-Poly & \aSuc & \aSuc & \aSuc & - & - & - \\
            C2 CBC-EtM &&&&&& \\
            \phantom{C2 } - \msgunknown & 0.0300 & 0.0003 & 0.0003 & - & - & - \\
            \phantom{C2 } - \msgping & 0.8383 & - & 0.0074 & - & - & -
            \vspace{0.3em}
            \\
            C3 Rogue Extension & \aNot & \aNot & \aNot & \aNot & \aSuc & \aNot
            \vspace{0.3em}
            \\
            C4 Rogue Session & \aNot & \aNot & \aNot & \aNot & \aSuc & \aNot
            \\
        \bottomrule
        \end{tabularx}
        \begin{tablenotes}
            \item[\aNot] Not evaluated.
            \item[\aSuc] Attack succeeds.
            \item[\aKlR] Client terminates the connection (rollover).
            \item[\aKlT] Client terminates the connection (timeout).
            \item[\aKlU] Client terminates the connection (\msgunknown message).
        \end{tablenotes}
        \caption{Expected outcomes for attacks against clients}
        \label{tbl:outcomes}
    \end{threeparttable}
\end{table}

\begin{compactdesc}
    \item[(C1):] \textbf{Sequence Number Manipulation (Sect.~4.1)}.
    \emph{We verified all techniques successfully against PuTTY~0.79.
    Additionally, our experiments show that OpenSSH~9.5p1 recognizes a rollover
    of sequence numbers and terminates the connection, thus not affected by any
    technique but \atkrcvinc. AsyncSSH~2.13.2 and libssh~0.10.5 allow for
    \atkrcvinc but terminate the connection due to handshake timeouts before
    any advanced technique concludes. Dropbear~2022.83 disconnects on
    \msgunknown messages instead of responding with \msgunimplemented but
    allows \xrcv to roll over, therefore being affected by \atkrcvinc and
    \atkrcvdec only.}

    \item[(C2):] \textbf{Extension Downgrade (Sect.~5.2)}. \emph{We
    successfully evaluated the attack in 10,000 trials on ChaCha20-Poly1305 and
    CBC-EtM against OpenSSH~9.5p1 and PuTTY~0.79 clients, connecting to
    OpenSSH~9.4p1 (\msgunknown only) and 9.5p1. For CBC-EtM, our success rate
    in practice was 0.0003 (OpenSSH) resp. 0.0300 (PuTTY), improved to 0.0074
    (OpenSSH) resp. 0.8383 (PuTTY) when sending \msgping instead of
    \msgunknown.}

    \item[(C3):] \textbf{Rogue Extension Negotiation (Sect.~6.1)}. \emph{We
    successfully evaluated the attack against AsyncSSH 2.13.2 as a client,
    connecting to AsyncSSH 2.13.2.}

    \item[(C4):] \textbf{Rogue Session Attack (Sect.~6.2)}. \emph{We
    successfully evaluated the attack against AsyncSSH 2.13.2 as a server,
    connecting to AsyncSSH 2.13.2.}
\end{compactdesc}

\paragraph{Internet Scan (Sect.~7)}
We are also including sample data, aggregated data, and evaluation scripts on
the Internet scan.

\subsubsection{Experiments}
The evaluation scripts (in the directory \verb|scripts|) are interactive and
self-describing. Some of them have several output files. In that case, the
files (as described below) are all opened in the text file viewer \verb|less|
\emph{at the same time}, requiring keyboard-based navigation to see all of the
results. As a gentle introduction to \verb|less|, see~\autoref{tbl:less} for a
quick reference of useful keyboard shortcuts.

\begin{table}
    \centering
    \begin{tabular}{ll}
        \toprule
        \textbf{Shortcut} & \textbf{Description} \\
        \midrule
        \textbf{q} & Quit \\
        \textbf{h} & Help \\
        \textbf{S} & Wrap long lines on/off \\
        \textbf{/} & Search \\
        \textbf{:n} & Next file \\
        \textbf{:p} & Previous file \\
        \bottomrule
    \end{tabular}
    \caption{Common keyboard shortcuts of \texttt{less}.}
    \label{tbl:less}
\end{table}

\begin{compactdesc}
    \item[(E1):] \texttt{test-sqn-manipulation.sh} [$\approx 1-3$ hours per
    client/variant combination]: Run one of the four sequence number
    manipulation attacks to prove \textbf{(C1)}. \atkrcvinc is very fast; the
    others can be slow.
    \begin{asparadesc}
        \item[Execution:] After starting the script, choose a client, one of
        the four attack options, and input the manipulation offset $N$. To
        prove (C1), input $N = 1$.
        \item[Results:] The attack is complete once the progress bar fills.
        After that, there will be an error message because the secure channel
        is broken, as the script does not implement any prefix truncation to
        complete the attack.
    \end{asparadesc}

    \item[(E2a):] \texttt{test-ext-downgrade.sh} [$\approx 1$ minute]: Run the
    extension downgrade attack to prove \textbf{(C2) for ChaCha20-Poly1305}.
    \begin{asparadesc}
        \item[Execution:] After starting the script, choose an arbitrary client
        and server combination. Afterward, choose attack variant \verb|1| to
        select ChaCha20-Poly1305.
        \item[Results:] The script will conclude by opening the following files
        simultaneously in \texttt{less}:
        \begin{enumerate}
            \item \texttt{diff} of files 3 and 4
            \item \texttt{diff} of files 5 and 6
            \item Server log (unmodified connection)
            \item Server log (tampered connection)
            \item Client log (unmodified connection)
            \item Client log (tampered connection)
            \item PoC proxy log
        \end{enumerate}
        Navigate to the second file. The file compares the output of the
        selected SSH client in the case of an extension downgrade attack to the
        output of an unmodified connection. The diff will indicate the presence
        of \texttt{SSH\_MSG\_EXT\_INFO} and absence of
        \texttt{SSH\_MSG\_IGNORE} in the unmodified connection only, thus
        proving (C2) for ChaCha20-Poly1305.
    \end{asparadesc}

    \item[(E2b):] \texttt{bench-ext-downgrade.sh} [$\approx 1-2$ hours per
    client/variant combination]: Run the extension downgrade attack 10,000
    times to prove \textbf{(C2) for CBC-EtM (\msgunknown and \msgping)}.
    \begin{asparadesc}
        \item[Execution:] After starting the script, choose between \msgunknown
        and \msgping variants of the attack, then select between OpenSSH and
        PuTTY client. A progress bar will show the current trial.
        \item[Results:] After finishing all trial connections, the number of
        successful trial runs will be outputted to the console. The relative
        success rate will be close to the values claimed in (C2), thus proving
        the functionality and success probability claims in (C2) in the case of
        CBC-EtM.
    \end{asparadesc}

    \item[(E3):] \texttt{test-asyncssh-rogue-ext-negotiation.sh} [$\approx 1$
    minute]: Run rogue extension attack to prove \textbf{(C3)}.
    \begin{asparadesc}
        \item[Execution:] The attack is automatic.
        \item[Results:] The script will conclude by opening a set of seven
        files in \texttt{less}. Refer to the results of (E2a) for a list of
        files opened. Navigate to the second file. The diff will indicate the
        presence of the \texttt{server-sig-algs} extension with an
        attacker-chosen value in the tampered connection, thus proving (C3).
    \end{asparadesc}

    \item[(E4):] \texttt{test-asyncssh-rogue-session-attack.sh} [$\approx 1$
    minute]: Run the rogue session attack to prove \textbf{(C4)}.
    \begin{asparadesc}
        \item[Execution:] The attack is automatic.
        \item[Results:] The script will conclude by opening a set of seven
        files in \texttt{less}. Refer to the results of (E2a) for a list of
        files opened. Navigate to the first file. The diff will indicate
        successful authentication for the victim (unmodified connection) and
        attacker (tampered connection), respectively. Afterward, navigate to
        the second file and examine the output of each client connection at the
        end of the file. In the unmodified connection, the server will respond
        with the username victim, while in the attacked connection, the server
        will respond with the username attacker. This proves (C4).
    \end{asparadesc}

    \item[(E5):] \texttt{scan\_util.py}  [$\approx 1$ minute]: Run the script
    to aggregate a set of zgrab2 scan results. Note that this script is in the
    sub-directory \verb|scans|. Without the full data, we can not prove the
    statistics in our paper. However, we can demonstrate how we aggregated the
    scan results and classified the algorithms.
    \begin{asparadesc}
        \item[Execution:] Build the docker image by running the following
        command inside the \verb|scans| directory:
\begin{lstlisting}
$ docker build . -t terrapin-artifacts/scan-util
\end{lstlisting}
        Now aggregate the \verb|sample.json| file which can be found in the
        \verb|sample| sub-directory by running the following command inside the
        \verb|scans| directory:
\begin{lstlisting}
$ docker run --rm -v ./sample:/input terrapin-artifacts/scan-util evaluate -i /input/sample.json -o /input/sample-ae.acc.json
\end{lstlisting}
        \item[Results:] The aggregation result will become available as
        \verb|sample-ae.acc.json| inside the \verb|sample| sub-directory. The
        total number of clients, status and version distribution, offered key
        exchange algorithms, comment strings, and other evaluation criteria
        will match the data present within the \verb|sample.json| file. Also,
        there is no difference between the \verb|sample-ae.acc.json| and
        \verb|sample.acc.json| files (aside from the evaluation start and end
        timestamps).
    \end{asparadesc}
\end{compactdesc}

\subsubsection{Troubleshooting}

\paragraph{Address already in use.}
If an attack script is interrupted, some docker containers may not be cleaned
up properly, blocking the server port permanently or for the duration of
TIME-WAIT (1~min. on Linux, 30~sec. on MacOS). 

Please follow these steps in this case:

\begin{enumerate}
    \item Run the script \verb|cleanup-system.sh|. This will stop and remove
    any pending Docker containers.

    \item If the problem persists, wait for up to 4~minutes.
\end{enumerate}

\paragraph{System Reset.}
To fully clean up the Docker containers and images, you can run
\verb|cleanup-system.sh --full|.

\subsection{Notes on Reusability}
The proof-of-concept code (\verb|pocs/|) has been kept short for simplicity and
is thus not modularized for reusability. However, the artifacts may serve as a
template for other MitM attacks on network protocols like SSH.

The Docker files for the evaluated SSH implementations (\verb|impl/|) may be
generally useful in other research on SSH.

Improvements to Wireshark for better dissection of SSH protocols (not included
in these artifacts) have been submitted and accepted upstream and will be
available in a future version of Wireshark.


\subsection{Version}
Based on the LaTeX template for Artifact Evaluation V20231005. Submission,
reviewing and badging methodology followed for the evaluation of this artifact
can be found at \url{https://secartifacts.github.io/usenixsec2024/}.